\documentclass[aps,prx,reprint,twocolumn, amsmath,amssymb,amsfonts,footinbib,
superscriptaddress,longbibliography,
notitlepage]{revtex4-1}

\AtBeginDocument{\usepackage{booktabs}}               
\makeatletter
\g@addto@macro\bfseries{\boldmath}
\makeatother

\usepackage[varg]{txfonts}
\usepackage[T1]{fontenc}
\usepackage[utf8]{inputenc}

\usepackage[varg]{txfonts}
\usepackage{hyperref}
\usepackage{amsmath}
\usepackage{color}
\usepackage{graphicx}
\usepackage[percent]{overpic}
\usepackage{mathrsfs}
\usepackage{bm}
\usepackage{braket}
\usepackage{mathtools}
\usepackage{bbold}
\usepackage{enumitem}

\usepackage{makecell}

\definecolor{cred}{RGB}{228,26,28}
\definecolor{cblue}{RGB}{8,48,107}
\definecolor{cgreen}{RGB}{77,175,74}
\definecolor{cgray}{RGB}{150,150,150}
\definecolor{clgray}{RGB}{200,200,200}
\definecolor{cpurple}{RGB}{152,78,163}
\definecolor{corange}{RGB}{255,127,0}
\definecolor{cgold}{RGB}{230,171,2}

\newcommand{\avg}[1]{\langle #1 \rangle}

\newcommand{\h}[1]{{#1}^{\dagger}} 
\newcommand{\cc}[1]{{#1}^{*}}

\newcommand{\subref}[2]{\ref{#1}\hyperref[#1]{#2}}

\renewcommand{\vec}[1]{\boldsymbol{#1}}
\renewcommand{\l}[1]{{#1}^{\phantom{\dagger}}} 
\newcommand{\mat}[1]{\vec{#1}}
\newcommand{\trp}[1]{{#1}^{\intercal}}
\newcommand{\vhat}[1]{\vec{\hat{#1}}}

\definecolor{cred}{RGB}{188,55,84}
\hypersetup{colorlinks=true,linkcolor=cred,citecolor=cred,urlcolor=cred}

\begin{document}

\title{Ground state properties of the Heisenberg-compass model on the square lattice}
\author{Subhankar Khatua}
\affiliation{Department of Physics, University of Windsor, 401 Sunset Avenue, Windsor, Ontario, N9B 3P4, Canada}
\affiliation{Department of Physics and Astronomy, University of Waterloo, Waterloo, Ontario, N2L 3G1, Canada}
\author{Griffin C. Howson}
\affiliation{Department of Physics, University of Windsor, 401 Sunset Avenue, Windsor, Ontario, N9B 3P4, Canada}
\affiliation{Department of Physics and Astronomy, University of Waterloo, Waterloo, Ontario, N2L 3G1, Canada}
\author{Michel J. P. Gingras}
\affiliation{Department of Physics and Astronomy, University of Waterloo, Waterloo, Ontario, N2L 3G1, Canada}
\author{Jeffrey G. Rau}
\affiliation{Department of Physics, University of Windsor, 401 Sunset Avenue, Windsor, Ontario, N9B 3P4, Canada}

\begin{abstract}
Compass models provide insights into the properties of Mott-insulating materials that host bond-dependent anisotropic interactions between their pseudospin degrees of freedom. 
In this article, we explore the classical and quantum ground state properties of one such model relevant to certain layered perovskite materials akin to Ba$_2$IrO$_4$ -- namely, the Heisenberg-compass model on the square lattice. 
We first investigate the ground state phase diagram of this model using classical Monte Carlo simulations. 
These reveal that the low temperature classical phase diagram is divided into six different classes of long-range ordered phases, including four phases that exhibit an order by disorder selection and two phases that are stabilized energetically. This model admits a special duality transformation, known as the Klein duality, conveniently allowing to map one region of coupling parameters onto another and constraining the phase diagram, and which we exploit in our study. 
From the analysis of the zero-point energy and the free energy of the spin waves, we find that order by quantum disorder at zero temperature and order by thermal disorder select the same orderings as those found from classical Monte Carlo simulations. 
We further investigate the quantum ground states of this model using numerical exact diagonalization on small clusters by exploiting the translational symmetry of the square lattice. We obtain a ground-state phase diagram bearing close resemblance to that found from the classical analysis. 

\end{abstract}

\date{\today}

\maketitle
\section{Introduction}

Lattice models with effective spin-spin interactions offer a fundamental framework for understanding the intriguing behavior of Mott insulating magnets~\cite{Auerbach1998,Fazekas1999}. A notable subclass of these models is the family of compass models that have gained significant attention in the context of strongly correlated transition metal (TM) oxides~\cite{Nussinov-compass-rev}. These models are characterized by spatially direction-dependent interactions among their spins. 
The details of such bond-dependent interactions depend strongly on the symmetries of the underlying system.
A well-studied compass model is the Kitaev spin model on the honeycomb lattice~\cite{Kitaev2006} which involves three Ising-like spin-spin compass couplings, namely $\cramped{x-x}$, $\cramped{y-y}$, or $\cramped{z-z}$, depending upon the bond orientations of the underlying lattice. 
This is in contrast to the isotropic Heisenberg interaction or anisotropic interactions that take the same form on every bond in the lattice. 
Compass interactions are often competing in nature or frustrated, which can lead to a continuous accidental classical ground state degeneracy that is not due to any symmetry of the Hamiltonian. 
Typically, such degeneracy is \emph{not} robust to thermal or quantum fluctuations at low energies, and is consequently lifted by these fluctuations, resulting in the stabilization of a magnetically long-range ordered state. 
This phenomenon is referred to as ``order by disorder''~\cite{Villain1980,Shender1982,Henley1989}. 
Conversely, in certain compass models, frustration can prevent ordering down to absolute zero temperature, giving rise to unconventional phases like spin liquids, for example in Kitaev and generalized Kitaev models~\cite{Kitaev2006,Mandal2009, Hermanns2014,Kevin2016,Eschmann2020}. Compass models thus provide a natural platform for exploring the interplay between conventional notions of order and more exotic disordered states that can arise in frustrated magnets.  

Compass models have garnered significant interest due to their applicability across various domains of condensed matter physics~\cite{Nussinov-compass-rev}. 
Initially, they were introduced to elucidate a range of physical phenomena in insulating TM oxides with weak spin-orbit coupling, wherein the orbital degrees of freedom of the TM ions couple with each other via compass interactions~\cite{Kugel1982, Feiner1997,Tokura2000,Khomskii2003, Brink2004, Mishra2004}. 
Over time, compass interactions have also been identified in materials with large spin-orbit coupling where the interactions are not between the orbitals, but rather between pseudospin degrees of freedom~\cite{Jackeli2009}. 
For instance, there has been a surge of efforts in realizing the Kitaev model in real materials with $4d$ and $5d$ TM ions, including $\alpha\textrm{-RuCl}_3$ and iridates~\cite{Jackeli2009,Rau2016,Takagi2019,Wang2017}. 
Compass models have also been found relevant in other contexts, such as $p+ip$ superconducting Josephson-junction arrays~\cite{Xu2004,Nussinov2005}, ultracold atoms trapped in optical lattices~\cite{Wu2008}, settings to safeguard qubits against unwanted decoherence in quantum computing~\cite{Doucot2005}, higher-form subsystem symmetry breaking~\cite{Rayhaun2023}, dimensional reduction~\cite{Batista2005}, and strongly interacting topological insulators~\cite{Tirrito2022}. 

Further, exploring material manifestations of compass interactions, individual layers of certain iridium-based perovskites, such as Ba$_2$IrO$_4$ and Tb-substituted Sr$_2$IrO$_4$, offer potential realizations of compass interactions on the square lattice~\cite{Katukuri2014,Zhang2016,Bertinshaw2019}.  
For example, the Hamiltonian of a single layer in Ba$_2$IrO$_4$ is thought to possess dominant antiferromagnetic Heisenberg and sub-dominant compass exchange~\cite{Katukuri2014}. Unlike in its cousin Sr$_2$IrO$_4$~\cite{Jackeli2009}, the IrO$_4$ octahedra in Ba$_2$IrO$_4$ do not undergo a staggered rotation~\cite{Okabe2011} and the Dzyaloshinskii-Moriya (DM) interaction is forbidden. 
While Ref.~[\onlinecite{Katukuri2014}] has provided a detailed analysis of a microscopic model of Ba$_2$IrO$_4$, the complete ground state phase diagram of the in-plane Hamiltonian as a function of Heisenberg and compass interactions has, to the best of our knowledge, not yet been explored. 
Note, however, that previous theoretical studies have considered more general Heisenberg-compass models on the square lattice~\cite{Trousselet2010,Trousselet2012}, wherein the compass couplings on the $x$ (horizontal) and $y$ (vertical) bonds of the square lattice were not constrained to be identical.
These investigations have revealed a rich quantum phase diagram with multiple phase transitions. On the other hand, the symmetric compass interaction limit (equal compass coupling on the $x$ and $y$ bonds) in the presence of isotropic Heisenberg exchange, which is relevant to the Ba$_2$IrO$_4$ layers, has not been examined in detail. 
Thus, investigating the Heisenberg-compass model on the square lattice with symmetric compass interactions is crucial to provide a foundation for understanding the behavior of these perovskites as well as similar spin-orbit coupled magnets on the square lattice that may be synthesized in the future. 

In this article, we determine the low-temperature classical and quantum phases of the symmetric compass model on the square lattice in the presence of an additional Heisenberg exchange interaction, hereafter referred to as the ``Heisenberg-compass model''. Using classical Monte Carlo simulations and spin-wave analysis, we first establish the classical phase diagram of this model. Our analysis reveals six distinct regimes in the Heisenberg-compass coupling-parameter space, each corresponding to a magnetically ordered phase. 
These phases are pair-wise related via a unitary transformation -- the so-called ``Klein duality'' familiar from generalized Kitaev models~\cite{Kimchi2014}.
Of particular interest to the present work are four long-range ordered phases that we identify as arising from order by disorder (ObD) induced by thermal fluctuations, while the remaining two result from conventional energetic selection. 
Next, we consider the quantum Heisenberg-compass model.
We find that in the semi-classical large spin limit, the free energy of quantum spin waves predicts a quantum ObD selection at zero temperature that matches the one found from classical thermal ObD in all four revelant coupling parameter regimes.
This persists to finite temperature, with combined quantum and thermal fluctuations preferring the same states at low non-zero temperatures.
Finally, we tackle the $S=1/2$ limit relevant for real materials.
Using exact diagonalization, we determine the quantum ground states at zero temperature, obtaining a qualitatively similar phase diagram to the classical and semi-classical limit, including the selection of the same states as the ones found from semi-classical quantum ObD.

\section{Model}
\label{sec.model}
\begin{figure}
\includegraphics[width=\columnwidth]{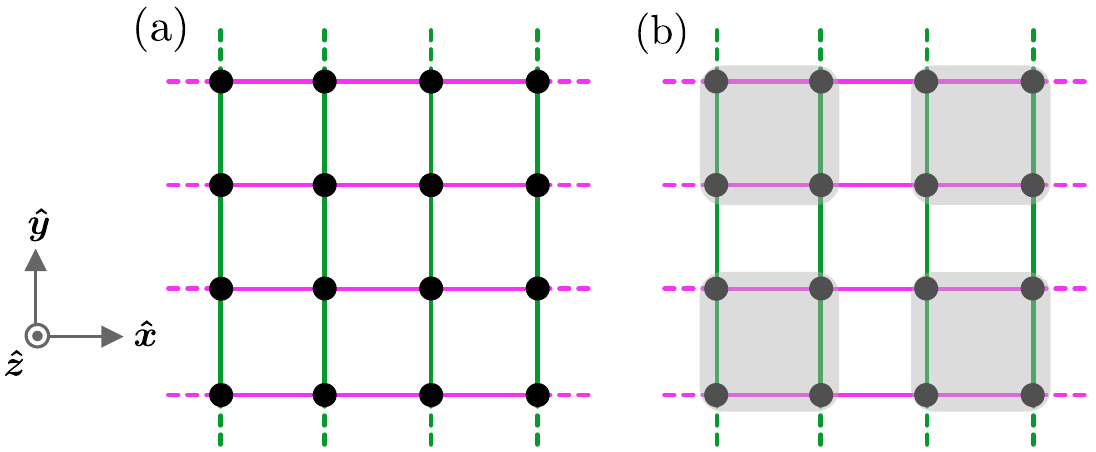}
\caption{The Heisenberg-compass model on the square lattice. (a) Each nearest-neighbor bond corresponds to an isotropic Heisenberg coupling, $J$, between the spins connected by the bond. Furthermore, there is a bond-dependent compass coupling, $K$, operating between the $x$ components of spins connected by the horizontal (magenta) bonds and between the $y$ components of spins connected by the vertical (green) bonds. (b) The square lattice is schematically shown with (grey) clusters of four spins. The Klein duality transformation of Eq.~\ref{eq.klein} applies identically on each cluster.}
\label{fig.model}
\end{figure}

We consider the Heisenberg-compass model on the square lattice defined by the Hamiltonian
\begin{equation}
   \label{eq.Ham}
  {\cal H} = \sum_{\vec{r}}\biggl[J\!\!\sum_{\vec{\delta} = \vec{x},\vec{y}} 
    \vec{S}^{\phantom{x}}_{\vec{r}} \!\cdot \vec{S}^{\phantom{x}}_{\vec{r}+\vec{\delta}\phantom{+\vec{}}}
    \!\!+K \left(S^x_{\vec{r}\phantom{+\vec{}}}\!\! S^x_{\vec{r}+\vec{x}} + S^y_{\vec{r}\phantom{+\vec{}}} \!\!S^y_{\vec{r}+\vec{y}} 
  \right)\biggr],
\end{equation}
where $\vec{S}^{\phantom{x}}_{\vec{r}}\equiv (S^x_{\vec{r}}, S^y_{\vec{r}},S^z_{\vec{r}})$ is a quantum spin-$1/2$ operator at  site $\vec{r}$, and $\vec{\delta} = \vec{x},\vec{y}$ denotes the nearest-neighbor (horizontal and vertical) bond with isotropic Heisenberg and anisotropic compass couplings $J$ and $K$, respectively, as shown in Fig.~\subref{fig.model}{(a)}. 
Since there are two coupling parameters $(J,K)$ in Eq.~\eqref{eq.Ham}, we parameterize them using an angle $\xi\in[0,2\pi)$ with $J \equiv \cos\xi$ and $K \equiv \sin\xi$, setting $\sqrt{J^2 + K^2}\equiv 1$ as the unit of energy with $\hbar\equiv k_{\rm B}\equiv1$. 
In other words, the interaction strengths are taken in such a way that they live on a circle of unit radius, as shown in Fig.~\ref{fig.phase-diagram}. While the dynamics of ObD from thermal fluctuations in the \emph{classical} ferromagnetic Heisenberg-compass model has recently been discussed~\cite{Khatua2023}, the full phase diagram of this model has not yet been explored.

There are a number of important and well-understood limits in the phase diagram of the model.
First, at $\xi = 0$ and $\xi = \pi$, the Hamiltonian [Eq.~\eqref{eq.Ham}] reduces to the well-known Heisenberg antiferromagnet $(J = 1)$ and ferromagnet $(J = -1)$, respectively, and marked by `HAF' and `HF' in Fig.~\ref{fig.phase-diagram}.
At these two special points, the model possesses a global SU(2) symmetry.
However, away from these points with a non-zero compass term $(K\neq 0)$, the Hamiltonian no longer has any continuous spin-rotation symmetry.
Nevertheless, it still possesses a discrete $C_4$ symmetry about the $\vhat{z}$ axis, which lies normal to the $\vhat{x}-\vhat{y}$ plane, and $C_2$ symmetries about the $\vhat{x}$ and $\vhat{y}$ axes. 
In the special cases of $\xi = \pi/2$ and $3\pi/2$, one has the well-studied antiferromagnetic and ferromagnetic ``pure'' compass model, respectively~\cite{Dorier2005,Mishra2004,Wenzel2008,Nussinov-compass-rev}. 
These two special points are related to one another by the following symmetry; a $\pi$-rotation of the spins about the $\vhat{z}$ axis on one of the two sublattices of the square lattice maps $K \rightarrow -K$.
We therefore refer to both of the ferromagnetic and antiferromagnetic compass models as `C' in Fig.~\ref{fig.phase-diagram}. At point C, extra discrete symmetries (special to that compass point) leads to a sub-extensive ground state degeneracy ($\sim 2^{L+1}$ for an $L\times L$ square lattice). In the classical limit at point C, in addition to this symmetry enforced sub-extensive ground state degeneracy, there are accidentally degenerate ground states forming a continuous $O(2)$ manifold~\cite{Dorier2005}. 
Interestingly, thermal or quantum fluctuations lift this accidental degeneracy, thus yielding an ObD of colinear states having long-range directional/nematic ordering along the director $\vhat{x}$ or $\vhat{y}$~\cite{Dorier2005}. 

\section{Klein duality}
\label{sec.klein-duality}
We now discuss the existence of a special unitary transformation within the Heisenberg-compass model [Eq.~\eqref{eq.Ham}] that strongly constrains the structure of its phase diagram. This transformation is the so-called Klein duality~\cite{Kimchi2014}, that maps one set of coupling parameters $(J, K)$ onto another set $(J',K')$. 
If the properties of the Heisenberg-compass model are known at $(J, K)$, then by using this transformation, one is able to determine its properties at $(J',K')$. 
This duality has been previously discussed in the context of Heisenberg-Kitaev models on the triangular~\cite{Khaliullin2005,Kimchi2014,Maksimov2019}, honeycomb~\cite{Kimchi2014,Chaloupka2015,Rau2016,Rousochatzakis2016}, kagome~\cite{Kimchi2014}, hyperkagome~\cite{Kimchi2014}, cubic ~\cite{Khaliullin2002,Khaliullin2003,Khaliullin2005}, FCC~\cite{Kimchi2014},
and pyrochlore lattices~\cite{Kimchi2014}.

This transformation consists of a four sublattice operation;  we divide the square lattice into clusters of four spins, as shown in Fig.~\subref{fig.model}{(b)} and apply the following transformation identically on each of the clusters:
\begin{align}
 \vec{S}'_{\vec{r}}&\equiv \mathbb{1}\,\vec{S}_{\vec{r}} = (S^x_{\vec{r}}, S^y_{\vec{r}},S^z_{\vec{r}}),\nonumber\\
\vec{S}'_{\vec{r}+\vec{x}}&\equiv \mathbb{R}^\pi_{\vhat{x}}\, \vec{S}_{\vec{r}+\vec{x}} = (S^x_{\vec{r}+\vec{x}}, \,-S^y_{\vec{r}+\vec{x}},\,-S^z_{\vec{r} +\vec{x}}),\nonumber\\
 \vec{S}'_{\vec{r}+\vec{y}}&\equiv \mathbb{R}^\pi_{\vhat{y}}\,\vec{S}_{\vec{r}+\vec{y}} = (-S^x_{\vec{r}+\vec{y}}, \,S^y_{\vec{r}+\vec{y}},\,-S^z_{\vec{r} +\vec{y}}),\nonumber\\
 \vec{S}'_{\vec{r}+\vec{x}+\vec{y}}&\equiv \mathbb{R}^\pi_{\vhat{z}}\,\vec{S}_{\vec{r}+\vec{x}+\vec{y}} = (-S^x_{\vec{r}+\vec{x}+\vec{y}}, \,-S^y_{\vec{r}+\vec{x}+\vec{y}},\,S^z_{\vec{r} +\vec{x}+\vec{y}}),
\label{eq.klein}
\end{align}
where $\mathbb{1}$ is the identity rotation and $\mathbb{R}^\pi_{\vhat{x}}$, $\mathbb{R}^\pi_{\vhat{y}}$, and $\mathbb{R}^\pi_{\vhat{z}}$ denote $\pi$-rotations about the $\vhat{x}$, $\vhat{y}$, and $\vhat{z}$ axes, respectively. 
The rotations, $\Big\{\mathbb{1}, \mathbb{R}^\pi_{\vhat{x}}, \mathbb{R}^\pi_{\vhat{y}}, \mathbb{R}^\pi_{\vhat{z}}\Big\}$ form an Abelian group, isomorphic to $\mathbb{Z}_2\times\mathbb{Z}_2$, known as the Klein four-group~\cite{humphreys1996}. 
Hence, the transformation in Eq.~\eqref{eq.klein} is referred to as the Klein transformation. 
Under this transformation, the Hamiltonian [Eq.~\eqref{eq.Ham}] maps back to itself, but with modified coupling parameters: $J\rightarrow J'\equiv -J$ and $K\rightarrow K'\equiv (2J+K)$. 
The mapping of the coupling parameters takes the following form,
\begin{equation}
  (J,K) \rightarrow (J',K')\equiv\frac{1}{\sqrt{J^2 + (2J+K)^2}}\Big(-J, 2J+K\Big),
  \label{eq.klein_duality}
\end{equation}
where the prefactor $1/\sqrt{J^2 +(2J+ K)^2}$ takes into account the renormalization of the overall energy scale from the parameter change in $(J,K) \rightarrow (J',K')$.
This transformation exactly maps the properties of the Heisenberg-compass model from one region of the $(J,K)$ parameter space
to another one, allowing us to restrict our focus to a smaller subset of the phase diagram and recover the properties of the model within the remaining regions by an application of the duality. 

\begin{figure}
\includegraphics[width=\columnwidth]{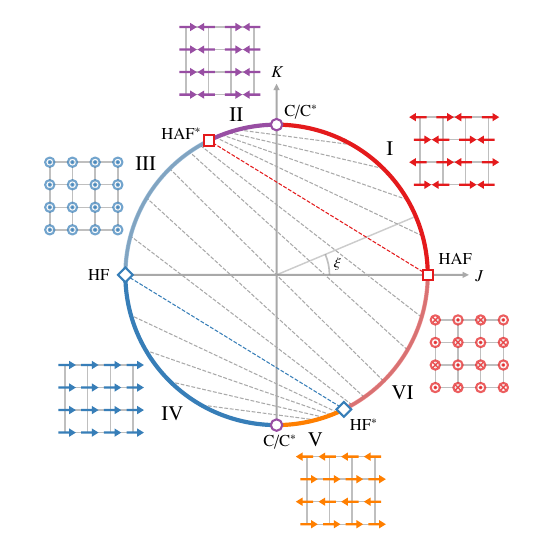}
\caption{Low temperature classical phase diagram of the Heisenberg-compass model. The whole parameter space is divided into six different regimes (labeled I through VI and marked by different colored arcs) separated by boundaries marked by different markers. Dashed lines connect the points on the circle that are related by the Klein duality. The points $J =+1 $ and $J = -1$ correspond to the Heisenberg antiferromagnet (HAF) and ferromagnet (HF), respectively. The $K = \pm 1$ points are the pure compass (C) limits, and by Klein duality, they are their self-duals (C = C*). The two pure compass limits ($\xi = \pi/2$ and $3\pi/2$) are also related by an exact symmetry (see text). The points, $\xi = \pi - \tan^{-1}(2)$ (marked as HAF*) and  $\xi = 2\pi - \tan^{-1}(2)$ (marked as HF*) are dual to HAF and HF, respectively. Six regimes exhibit six different ordering phases at low temperatures. To illustrate each phase, a corresponding representative configuration out of its several symmetry related copies is shown in the same color as that of its arc.  This phase diagram has been confirmed by performing classical Monte Carlo simulations down to low temperatures.}
\label{fig.phase-diagram}
\end{figure}

To visualize this mapping, we connect each pair of points in the parameter space that are dual to one another by dashed lines in Fig.~\ref{fig.phase-diagram}. 
We note from Eq.~\eqref{eq.klein_duality} that when $(2J+K) = 0$, the anisotropic Heisenberg-compass model maps onto the isotropic Heisenberg model $(K' = 0)$. 
We find that $(2J+K) = 0$ admits two solutions: $\xi =\pi -\tan^{-1}(2)$ and $2\pi - \tan^{-1}(2)$. As shown in Fig.~\ref{fig.phase-diagram}, $\xi =\pi -\tan^{-1}(2)$ is dual to the pure Heisenberg antiferromagnet (HAF) and labelled as HAF*. 
Similarly, $\xi =2\pi -\tan^{-1}(2)$ is dual to the pure Heisenberg ferromagnet (HF) and marked as HF*. 
The Klein duality also reveals that the pure compass points ($\xi = \pi/2, 3\pi/2)$ are their self-dual, marked as C*, i.e., they map back to themselves (see Fig.~\ref{fig.phase-diagram}). 

The entire parameter space is naturally divided into six regimes; Regime-I (HAF -- C), Regime-II (C -- HAF*), Regime-III (HAF* -- HF), Regime-IV (HF -- C), Regime-V (C -- HF*), and Regime-VI (HF* -- HAF). 
Using the above duality mapping, as depicted in Fig.~\ref{fig.phase-diagram}, it is clear that Regime-II, V, and VI are dual to Regime-I, IV, and III, respectively. For simplicity, we shall restrict ourselves only to Regime-I, III, and IV, obtaining the properties of Regime-II, V and VI by exploiting the Klein duality.

\section{Classical ground states}
\label{sec.cgs}
We start by considering the model of Eq.~\eqref{eq.Ham}  \emph{classically} where $\vec{S}_{\vec{r}}$ is a three component vector of fixed length $S$ at site $\vec{r}$. For this classical model, we can consider a further transformation given by 
\begin{eqnarray}
    \vec{S}_{\vec{r}} \rightarrow (-1)^{\vec{r}}\,\vec{S}_{\vec{r}},
    \label{eq.sublattice-flip}
\end{eqnarray}
that flips all the spins on one of the two sublattices of the square lattice. This sublattice spin-flip transformation keeps the classical Hamiltonian [Eq.~\eqref{eq.Ham}] invariant if the coupling parameters $(J,K)$ are changed to $(-J,-K)$, i.e., $\xi\rightarrow \xi+\pi$. 
Similar to the Klein duality [Eq.~\eqref{eq.klein}], the transformation in Eq.~\eqref{eq.sublattice-flip} maps different regions of the coupling parameter space onto each other, specifically, relating Regimes I, II and III to Regimes IV, V, and VI, respectively.
However, it is important to note that reversing the sign of a spin is not a canonical transformation as the Poisson bracket relation for the spin components is not preserved: $\cramped{\{S^\mu_{\vec{r}}, S^\nu_{\vec{r}}\} = \epsilon_{\mu\nu\delta}S^\delta_{\vec{r}}}$ changes to $\cramped{\{S^\mu_{\vec{r}}, S^\nu_{\vec{r}}\} = -\epsilon_{\mu\nu\delta}S^\delta_{\vec{r}}}$. Consequently, not all properties of the model at the spin-flip transformation related parameter sets $(J, K)$ and $(-J,-K)$ are directly related.
For instance, the dynamical properties of the transformed Hamiltonian are different from that of the original Hamiltonian since the equations of motion are not preserved under the transformation of Eq.~\eqref{eq.sublattice-flip}.
This is in contrast to the canonical Klein duality transformation [Eq.~\eqref{eq.klein}] where \emph{all} model properties can be mapped exactly between two parameter sets. 
Nevertheless, as far as the thermodynamic properties of the classical model are concerned, the sublattice spin-flip transformation [Eq.~\eqref{eq.sublattice-flip}] provides an exact mapping between the parameter sets as the partition function remains unchanged under this transformation. This includes determining the classical ground states and the low-temperature phases of the model, thereby further constraining the classical phase diagram, in addition to the constraints imposed by the Klein duality. 
Consequently, for the analysis of the \emph{classical} phase diagram, we may focus on Regime-I and Regime-III, while employing the transformations of Eq.~\eqref{eq.klein} and Eq.~\eqref{eq.sublattice-flip} to determine the properties of the model in the remaining portions of the phase diagram. 
 
We begin our analysis using the Luttinger-Tisza method~\cite{Kimchi2014,Luttinger1946,Niggemann2020} to obtain the candidate classical ground states for a given phase angle $\xi$. 
The details of the calculations are presented in Appendix~\ref{app.LT}. 
We find that the Luttinger-Tisza method yields (spin length) normalizable states across the full phase diagram, providing the exact classical ground states for all values of $\xi$. 
\begin{enumerate}[label=(\roman*),leftmargin=*]
\itemsep0em 
\item \emph{Regime-I}: The ground states consist of N\'eel configurations in the $\vhat{x}-\vhat{y}$ plane with an arbitrary N\'eel direction. These configurations are given by
\begin{equation}
\vec{S}_{\vec{r}} = (-1)^{\vec{r}}S(\cos\phi\,\vhat{x} + \sin\phi\,\vhat{y}),
\label{eq.neel}
\end{equation}
where $\phi\in[0,\,2\pi)$. 
Thus, for any $\xi$ in Regime-I, the classical ground states form a continuous $O(2)$ manifold of in-plane N\'eel states parametrized by angle $\phi$. 
These ground states are \emph{accidentally} degenerate since continuous spin-rotations do not leave the anisotropic compass term of the Hamiltonian of Eq.~\eqref{eq.Ham} invariant. 

\item \emph{Regime-II}: Since this regime is dual to Regime-I, the ground states for any $\xi$ in Regime-II can be found from the N\'eel states [Eq.~\eqref{eq.neel}] using the Klein duality transformation in Eq.~\eqref{eq.klein}. 
These are given by the following four-site magnetic order, 
\begin{eqnarray}
    \vec{S}_{\vec{r}} &=& S(+\cos{\phi}\,\vhat{x}+\sin{\phi}\,\vhat{y}),\nonumber\\
    \vec{S}_{\vec{r}+\vec{x}} &=& S(-\cos{\phi}\,\vhat{x}+\sin{\phi}\,\vhat{y}),\nonumber\\
    \vec{S}_{\vec{r}+\vec{y}} &=& S(+\cos{\phi}\,\vhat{x}-\sin{\phi}\,\vhat{y}) = - \vec{S}_{\vec{r}+\vec{x}},\nonumber\\
    \vec{S}_{\vec{r}+\vec{x}+\vec{y}} &=& S(-\cos{\phi}\,\vhat{x}-\sin{\phi}\,\vhat{y}) = -\vec{S}_{\vec{r}}.
    \label{eq.case2}
\end{eqnarray}
As in Regime-I, these ground states are \emph{accidentally} degenerate as well, forming again an $O(2)$ manifold. These ground states become colinear stripe configurations for $\phi = 0,\,\pi/2,\,\pi,\,3\pi/2$ [see Eq.~\eqref{eq.case2}]. 
For example, the states with $\phi=0$ or $\pi$ can be described as ferromagnetically-ordered spins aligned parallel or antiparallel to $\vhat{x}$ within each column of the square lattice, with neighbouring columns ordered antiferromagnetically. Thus, for these states, the spins are either aligned along $\vhat{x}$ or $-\vhat{x}$, with the ordering wave vector or the `stripe direction' along $\vhat{x}$. 
On the other hand, the states for  $\phi = \pi/2$ or $3\pi/2$ are ferromagnetically-ordered rows of spins aligned along $\vhat{y}$ or $-\vhat{y}$, but arranged antiferromagnetically across neighboring rows. Here, the spins are oriented either along $\vhat{y}$ or $-\vhat{y}$, with the ordering wave vector or the stripe direction along $\vhat{y}$. Thus, these four stripe states have spins aligned either parallel or antiparallel to the stripe directions. Hereafter, we shall refer to these colinear stripe states [$\phi = 0,\,\pi/2,\,\pi,\,3\pi/2$ in Eq.~\eqref{eq.case2}] simply as ``Stripe-$\parallel$''. 

\item \emph{Regime-III}: We find that there are only two discrete ground states which correspond to ferromagnetic configurations along the $\pm\vhat{z}$ directions. 
Unlike Regime-I and II, these ground states are \emph{not} accidentally degenerate; they are related by an exact global $C_2$ symmetry about the $\vhat{x}$ or $\vhat{y}$ axes. 

\item \emph{Regime-IV}: Ground states in Regime-I and Regime-IV are related by the sublattice spin-flip transformation given in Eq.~\eqref{eq.sublattice-flip}. Specifically, the ground states in Regime-I, the N\'eel states in the $\vhat{x}-\vhat{y}$ plane, map to the uniform ferromagnetic configurations in the $\vhat{x}-\vhat{y}$ plane. Thus, the ground states in Regime-IV correspond to a ferromagnet with the magnetization along any arbitrary direction in the $\vhat{x}-\vhat{y}$ plane,
 \begin{equation}
\vec{S}_{\vec{r}} = S(\cos\phi\,\vhat{x} + \sin\phi\,\vhat{y}).
\label{eq.ferro}
\end{equation}
We thus have in Regime-IV an $O(2)$ manifold of \emph{accidentally} degenerate classical ground states, as we found in Regime-I. 

\item \emph{Regime-V}: This regime is Klein dual to Regime-IV with the classical ground states in Regime-V found from those in Regime-IV. Using the Klein duality transformation [Eq.~\eqref{eq.klein}] on the ferromagnetic states described in Eq.~\eqref{eq.ferro}, we obtain the following four site unit cell magnetic ordering for Regime-V,
\begin{eqnarray}
    \vec{S}_{\vec{r}} &=& S(+\cos{\phi}\,\vhat{x}+\sin{\phi}\,\vhat{y}),\nonumber\\
    \vec{S}_{\vec{r}+\vec{x}} &=& S(+\cos{\phi}\,\vhat{x}-\sin{\phi}\,\vhat{y}),\nonumber\\
    \vec{S}_{\vec{r}+\vec{y}} &=& S(-\cos{\phi}\,\vhat{x}+\sin{\phi}\,\vhat{y}) = - \vec{S}_{\vec{r}+\vec{x}},\nonumber\\
    \vec{S}_{\vec{r}+\vec{x}+\vec{y}} &=& S(-\cos{\phi}\,\vhat{x}-\sin{\phi}\,\vhat{y}) = -\vec{S}_{\vec{r}}.
    \label{eq.case5}
\end{eqnarray}

Being Klein dual to the $O(2)$ degenerate states of Regime-IV, these ground states are thus also \emph{accidentally} degenerate and form an $O(2)$ manifold. Equation~\eqref{eq.case5} reduces to colinear stripe states for $\phi = 0,\,\pi/2,\,\pi,\,3\pi/2$. 
The states with $\phi = 0, \pi$ have ferromagnetically-ordered spins aligned along $\vhat{x}$ or $-\vhat{x}$ within each row of the square lattice, with neighboring rows arranged antiferromagnetically and form stripes whose direction is along $\vhat{y}$. 
The states with $\phi = \pi/2, 3\pi/2$ have ferromagnetically-ordered spins aligned along $\vhat{y}$ or $-\vhat{y}$ within each column, but ordered antiferromagnetically across neighboring columns. These two configurations thus have a stripe direction that is along $\vhat{x}$. 
In contrast to the stripe states in Regime-II (Stripe-$\parallel$) where the spins are either parallel or antiparallel to the stripe directions, here in Regime-V, in the four stripe states, the spins are \emph{perpendicular} to the stripe directions. We shall refer to these colinear stripe states [$\phi = 0,\,\pi/2,\,\pi,\,3\pi/2$  in Eq.~\eqref{eq.case5}] as ``Stripe-$\perp$''. Note that these states can also be found by applying the sublattice spin-flip transformation in Eq.~\eqref{eq.sublattice-flip} on Stripe-$\parallel$. 

\item \emph{Regime-VI}: The ground states in this regime can be found from those of its dual Regime-III. 
The Klein duality transformation [Eq.~\eqref{eq.klein}] on the two ferromagnetic ground states along $\pm\vhat{z}$ directions [ground states in Regime-III] provides two discrete N\'eel states along the $\pm\vhat{z}$ directions. Again, these ground states can also be found from the ground states in Regime-III by applying the sublattice spin-flip transformation of Eq.~\eqref{eq.sublattice-flip}.
As with Regime-III, these N\'eel states are also \emph{not} accidentally degenerate, being related to one another by an exact global $C_2$ symmetry about the $\vhat{x}$ or $\vhat{y}$ axis. 
\end{enumerate}

To summarize, we have found the classical ground states of the Hamiltonian in Eq.~\eqref{eq.Ham} for all $\xi$ using the Luttinger-Tisza method, exploiting the Klein duality [Eq.~\eqref{eq.klein}] and the sublattice spin-flip [Eq.~\eqref{eq.sublattice-flip}] relation between various parameter regimes. 
For any value of $\xi$ in Regime-I, II, IV, and V, the ground states are accidentally degenerate, forming a continuous $O(2)$ manifold. 
However, for any $\xi$ in Regime-III and VI, we have only two discrete ground states that are related by an exact $C_2$ symmetry about the $\vhat{x}$ or $\vhat{y}$ axis. 
Six representative classical spin configurations corresponding to the six parameter regimes are shown in Fig.~\ref{fig.phase-diagram}. 
Since there exists a continuous accidental ground state degeneracy for four parameter regimes and this degeneracy is not a consequence of any exact symmetry of the Hamiltonian, one must next determine whether quantum or thermal fluctuations can lift this degeneracy through  ObD, as discussed in the Introduction.

\section{Low-temperature classical phase diagram}
\label{sec.phase-diagram}
We have discussed the classical ground states of the Heisenberg-compass model [Eq.~\eqref{eq.Ham}] in Sec.~\ref{sec.cgs} and shown that there exists accidental degeneracy in four of the parameter regimes. Given this, we next explore the question of ObD from thermal fluctuations in those regimes. In doing so, we determine the low-temperature classical phase diagram for the full parameter space.  

\subsection{Order by disorder from thermal fluctuations} 
\label{ssec.obtd}
To determine ObD from thermal fluctuations at low temperatures for a given $\xi$, we consider a classical spin-wave expansion about each of the ordered ground states for that $\xi$, and investigate how the spin-wave excitations contribute to the free energy of the system~\cite{Kawamura1984,McClarty2014,Seabra2016,Yan2017}. 
We assume that we are at sufficiently low temperature such that there are only small fluctuations (spin-wave excitations) about a particular classical ground state, i.e. a harmonic expansion is valid.  
The accidentally degenerate ground state for which the free energy of the harmonic spin wave fluctuations is minimal is the one selected by ObD at low temperatures.           
This (classical) free energy can alternatively be obtained from the \emph{quantum} non-interacting or linear spin-wave theory, as shown in Ref.~\cite{Seabra2016}. 
If, for a given $\xi$, the linear spin-wave spectrum at wave vector $\vec{q}$ about a classical ordered state parametrized by angle $\phi$ is $\omega_{\vec{q}}(\phi)$, the classical free energy is given by $\cramped{F(\phi) = T\sum_{\vec{q}}\ln\omega_{\vec{q}}(\phi)}$~\cite{McClarty2014,Seabra2016} where $\omega_{\vec{q}}(\phi)$ implicitly depends on $\xi$. 
A quantum linear spin-wave analysis using the Holstein-Primakoff formalism~\cite{Auerbach1998} is discussed in Appendix~\ref{app.SW-analysis}, deriving the frequencies of the linear spin-wave modes in the different parameter regimes. 
This alternative route to compute the free energy of the classical spin waves is convenient as we shall reuse the results of the quantum linear spin wave analysis in the context of ObD selection from quantum fluctuations in Sec.~\ref{ssec.qobd} and Sec.~\ref{ssec.qobd-thermal}.

\begin{table*}
\begin{center}
\begin{tabular}{c@{\hskip 0.15in}c@{\hskip 0.15in}c@{\hskip 0.15in}c@{\hskip 0.15in}c@{\hskip 0.15in}c} 
\midrule
\midrule
Regime & \makecell{Classical \\ ground states} &\makecell[c]{Relevant\\ magnetization} & \makecell[c]{Order\\ parameter}& \makecell{Low-temperature\\phase}&\makecell{Ordering\\ mechanism}\\ 
\midrule 
\addlinespace
I& \makecell[c]{$\vhat{x}-\vhat{y}$ N\'eel \\ \rm{[Eq.~\eqref{eq.neel}]}} &$\vec{m}_{\rm I} \equiv \frac{1}{N}\sum_{\vec{r}}(-1)^{m+n}\vec{S}_{\vec{r}}$ & $O_{\rm I} \equiv \sqrt{(m_{\rm I}^x)^2 + (m_{\rm I}^y)^2}\cos(4\phi_{\rm I})$ &\makecell[c]{N\'eel state \\along $\pm\vhat{x}$, $\pm\vhat{y}$} &ObD \\
\addlinespace
\addlinespace
II & \makecell[c]{States\\ in \rm{ Eq.~\eqref{eq.case2}}} & $\vec{m}_{\rm II} \equiv \frac{1}{N}\sum_{\vec{r}} \Big((-1)^m S^x_{\vec{r}}, (-1)^n S^y_{\vec{r}},0\Big)$& $O_{\rm II} \equiv \sqrt{(m_{\rm II}^x)^2 + (m_{\rm II}^y)^2}\cos(4\phi_{\rm II})$&\makecell[c]{Stripe-$\parallel$}&ObD \\
 \addlinespace
 \addlinespace
III & \makecell[c]{Ferromagnet \\along $\pm\vhat{z}$} &$\vec{m}_{\rm III} \equiv \frac{1}{N}\sum_{\vec{r}}\vec{S}_{\vec{r}}$& $O_{\rm III} \equiv |m_{\rm III}^z|$&\makecell[c]{Ferromagnet \\along $\pm\vhat{z}$} &Energetic \\
\addlinespace
\addlinespace
IV & \makecell[c]{$\vhat{x}-\vhat{y}$ ferromagnet\\ \rm{[Eq.~\eqref{eq.ferro}]}} &$\vec{m}_{\rm IV} \equiv \vec{m}_{\rm III}$ & $O_{\rm IV} \equiv \sqrt{(m_{\rm IV}^x)^2 + (m_{\rm IV}^y)^2}\cos(4\phi_{\rm IV})$  &\makecell[c]{Ferromagnet \\along $\pm\vhat{x}$, $\pm\vhat{y}$}&ObD \\
\addlinespace
\addlinespace
V & \makecell[c]{States\\ in \rm{Eq.~\eqref{eq.case5}}} &$\vec{m}_{\rm V} \equiv \frac{1}{N}\sum_{\vec{r}} \Big((-1)^n S^x_{\vec{r}}, (-1)^m S^y_{\vec{r}},0\Big)$& $O_{\rm V} \equiv \sqrt{(m_{\rm V}^x)^2 + (m_{\rm V}^y)^2}\cos(4\phi_{\rm V})$&\makecell[c]{Stripe-$\perp$}&ObD \\
\addlinespace
\addlinespace
VI & \makecell[c]{N\'eel state \\along $\pm\vhat{z}$}& $\vec{m}_{\rm VI} \equiv \vec{m}_{\rm I}$ & $O_{\rm VI} \equiv |m_{\rm VI}^z|$& \makecell[c]{N\'eel state \\along $\pm\vhat{z}$} & Energetic\\ 
\addlinespace
\midrule
\midrule
\end{tabular}
\caption{Summarizing the key results of the classical Heisenberg-compass model~[Eq.~\eqref{eq.Ham}]. We take $\vec{r}\equiv m\vhat{x} + n\vhat{y}$ where $m$ and $n$ assume values $0,1,\cdots , (L-1)$ for an $L\times L$ square lattice. The angle $\phi_{\rm i}$ in the order parameter is defined as: $\phi_{\rm i} \equiv \tan^{-1}(m_{\rm i}^y/m_{\rm i}^x)$ where i $=$ I, II, IV, V.} 
\label{tab.summary}
\end{center}
\end{table*}
\begin{figure*}
\includegraphics[width=\textwidth]{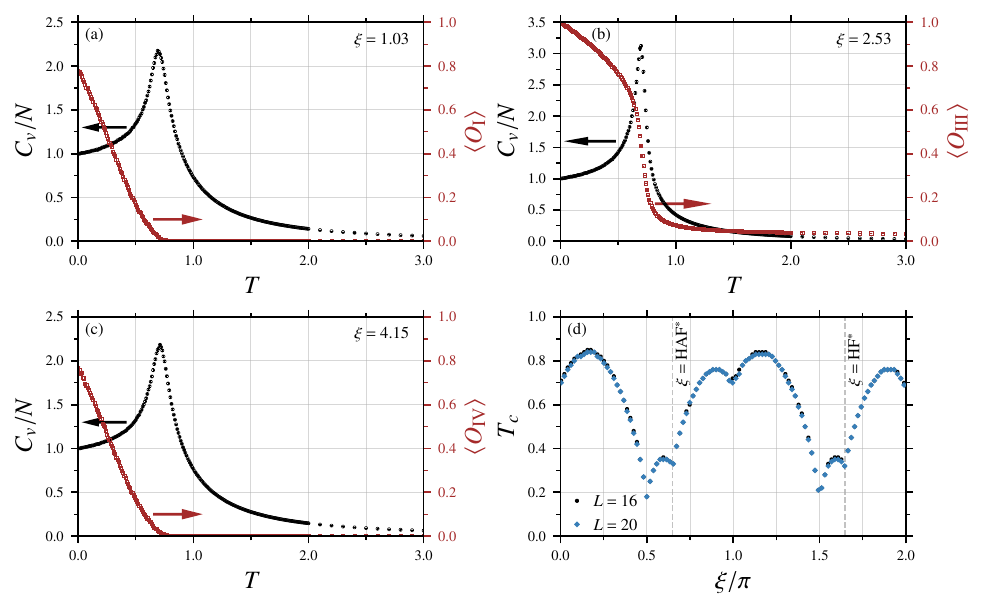}
\caption{Results obtained using classical Monte Carlo simulations. (a) Specific heat per spin, $C_v/N$, and the thermal average of the order parameter, $\langle O_{\rm I}\rangle $ vs temperature, $T$ for $\xi = 1.03$ in Regime-I, (b) $C_v/N$ vs $T$ and $\langle O_{\rm III}\rangle $ vs $T$ for $\xi = 2.53$ in Regime-III, and (c) $C_v/N$ vs $T$ and $\langle O_{\rm IV}\rangle$ vs $T$ for $\xi = 4.15$ in Regime-IV for $L = 16$. MC error bars on $C_v$ and the order parameters are found to be smaller than the marker size. (d) Critical temperature, $T_c$ vs the phase angle, $\xi$ found from the location of the peak of the specific heat for each $\xi$ for $L = 16$ and $L = 20$.}
\label{fig.sp-heat-ord-params}
\end{figure*}

In Regime-I, II, IV, and V, where there is an accidental $O(2)$ degeneracy among the classical ground states, the spin-wave free energy $F(\phi)$ depends on the ground state (parametrized by $\phi$)  about which the spin-wave analysis is performed. Minima are found only about certain discrete states out of the continuous $O(2)$ manifold of states in all of the above four regimes, a demonstration of ObD selection.
For any $\xi$ in Regime-I, the free energy is minimized over four N\'eel states along $\pm\vhat{x},\pm\vhat{y}$, corresponding to $\phi = 0,\pi/2,\pi, 3\pi/2$ in Eq.~\eqref{eq.neel}. 
Note that these four states are related to one another by a $C_4$ rotation about the $\vhat{z}$ axis, a consequence of the exact $C_4$ symmetry of the Hamiltonian in Eq.~\eqref{eq.Ham}.  
Now, applying the duality transformation [Eq.~\eqref{eq.klein}] to the states selected by ObD in Regime-I maps them to stripe states along $\pm\vhat{x}, \pm\vhat{y}$ directions corresponding to $\phi = 0, \pi/2,\pi,3\pi/2$ in Eq.~\eqref{eq.case2} (Stripe-$\parallel$), giving the corresponding ObD selected states for any $\xi$ in Regime-II. 
In Regime-III, the low-temperature classical phase is described by two discrete ground states -- the ferromagnetic states along $\pm\vhat{z}$ directions. 
The Klein duality between Regime-III and VI demands that the phase in Regime-VI is given by two N\'eel states along $\pm\vhat{z}$ directions.     
Finally, in Regime-IV, the classical spin-wave free-energy, $F(\phi)$, is minimized and gives ObD selection for the four ferromagnetic states along $\pm\vhat{x},\pm\vhat{y}$ directions corresponding to $\phi = 0,\pi/2,\pi,3\pi/2$ in Eq.~\eqref{eq.ferro}. 
By the Klein duality, in Regime-V, ObD selects four stripe states corresponding to $\phi = 0,\pi/2,\pi,3\pi/2$ in Eq.~\eqref{eq.case5} (Stripe-$\perp$). 
The above analysis determines the complete low-temperature classical phase diagram of the Heisenberg-compass model on the square lattice. 
A representative state of the symmetry related classical phases selected by ObD for each of the four parameter regimes with accidential degeneracy is illustrated beside each of the corresponding regimes in Fig.~\ref{fig.phase-diagram}.

\subsection{Classical Monte Carlo}
\label{ssec.MC}
To confirm the low-temperature classical phases found from the analysis of spin-wave free energy in Sec.~\ref{ssec.obtd}, we perform classical Monte Carlo (MC) simulations over a range of temperature spanning from well-below the ordering temperature, to well-above. 
To expose the orderings at low temperatures, we define the following order parameters for each of the six phases:
\begin{enumerate}[label=(\roman*),leftmargin=*]
\itemsep0em 
\item \emph{Regime-I}: The order parameter depends on the staggered magnetization $\vec{m}_{\rm I} \equiv (m_{\rm I}^x,m_{\rm I}^y,m_{\rm I}^z)\equiv (1/N)\sum_{\vec{r}}(-1)^{\vec{r}}\vec{S}_{\vec{r}}$. 
We define the order parameter, $O_{\rm I} \equiv \sqrt{(m_{\rm I}^x)^2 + (m_{\rm I}^y)^2}\cos(4\phi_{\rm I})$ where $\phi_{\rm I} \equiv \tan^{-1}(m_{\rm I}^y/m_{\rm I}^x)$, giving $O_{\rm I} = 1$ in the N\'eel states along the $\pm\vhat{x}, \pm\vhat{y}$ directions since $\sqrt{(m_{\rm I}^x)^2 + (m_{\rm I}^y)^2} = 1$ and $\cos(4\phi_{\rm I}) = 1$ for those states. 
However, if there were no selection of states at low temperatures and all the classical ground states were equally likely, then the thermal average of the order parameter, $\langle O_{\rm I}\rangle$ would vanish since the average of $\cos(4\phi_{\rm I})$ over the full range of $\phi_{\rm I}\in [0,2\pi)$ is zero. 

\item  \emph{Regime-II}: Here, we define a different magnetization, motivated from Eq.~\eqref{eq.case2}, $\vec{m}_{\rm II} \equiv \frac{1}{N}\sum_{\vec{r}} \Big((-1)^m S^x_{\vec{r}}, (-1)^n S^y_{\vec{r}},0\Big)$ with $\vec{r} = m\,\vec{x} + n\,\vec{y}$ and $m$, $n$ take values $0, 1, \cdots, (L-1)$ for an $L\times L$ square lattice. 
We then define the order parameter, $O_{\rm II} \equiv \sqrt{(m_{\rm II}^x)^2 + (m_{\rm II}^y)^2}\cos(4\phi_{\rm II})$ where $\phi_{\rm II} \equiv \tan^{-1}(m_{\rm II}^y/m_{\rm II}^x)$. 
Note that for ObD selected states in this regime, discussed in Sec.~\ref{ssec.obtd}, one has $O_{\rm II} = 1$. 

\item  \emph{Regime-III}: The order parameter is the absolute value of the $z$-component of the net magnetization, $O_{\rm III} \equiv |(1/N)\sum_{\vec{r}}S_{\vec{r}}^z|$, giving $O_{\rm III} = 1$ for ferromagnetic states aligned along the $\pm\vhat{z}$ directions. 

\item  \emph{Regime-IV}: We define the order parameter to be, $O_{\rm IV} \equiv \sqrt{(m_{\rm IV}^x)^2 + (m_{\rm IV}^y)^2}\cos(4\phi_{\rm IV})$, where $m_{\rm IV}^x$ and $m_{\rm IV}^y$ are respectively the $x$ and $y$ components of the net magnetization per spin and $\phi_{\rm IV} \equiv \tan^{-1}(m_{\rm IV}^y/m_{\rm IV}^x)$. 
The ObD selected states in this regime, i.e., the ferromagnetic configurations along $\pm\vhat{x},\pm\vhat{y}$ yields $\sqrt{(m_{\rm IV}^x)^2 + (m_{\rm IV}^y)^2} = 1 $ and $\cos(4\phi_{\rm IV}) = 1$, resulting in $O_{\rm IV} = 1$. 

\item  \emph{Regime-V}: Here, we define the order parameter in the same vein as we did for Regime-II. 
With the magnetization defined as $\vec{m}_{\rm V} \equiv \frac{1}{N}\sum_{\vec{r}} \Big((-1)^n S^x_{\vec{r}}, (-1)^m S^y_{\vec{r}},0\Big)$ with $\vec{r} = m\vhat{x} + n\vhat{y}$. 
The order parameter $O_{\rm V} \equiv \sqrt{(m_{\rm V}^x)^2 + (m_{\rm V}^y)^2}\cos(4\phi_{\rm V})$ where $\phi_{\rm V} \equiv \tan^{-1}(m_{\rm V}^y/m_{\rm V}^x)$, giving $O_{\rm V} = 1$ for the ObD selected states in this regime, discussed in Sec.~\ref{ssec.obtd}. 

\item  \emph{Regime-VI}: For this regime, the order parameter is the $z$ component of the staggered magnetization, $O_{\rm VI} \equiv |(1/N)\sum_{\vec{r}}(-1)^{\vec{r}}S_{\vec{r}}^z|$, characterizing the N\'eel states along the $\pm\vhat{z}$ directions.
\end{enumerate}

These order parameters along with the regime-relevant magnetizations are summarized in Table~\ref{tab.summary}.
Based on the low-temperature classical expansion described in Sec.~\ref{ssec.obtd}, we expect that the thermal average of the order parameters defined for a particular regime should approach unity in that regime (and approach zero elsewhere) as temperature is decreased towards zero. 
We measure the thermal averages of these various order parameters using classical MC simulations at low temperatures. 
The simulations are performed on a square lattice with $N = L^2$ sites assuming periodic boundary conditions. 
The spins under consideration are the three component (Heisenberg) spins of unit length (i.e. $|{\bm{S}}_{\bm{r}}|=1$).
The MC simulations are carried out based on an adaptive single-site Metropolis algorithm~\cite{Alzate-Cardona2019}, combined with over-relaxation moves~\cite{Creutz1987}. 
We define a Monte Carlo sweep at a certain temperature as a combination of adaptive single-site Metropolis moves successively at $N$ randomly chosen sites with each followed by over-relaxation moves consecutively at 5 randomly chosen sites. 

The full range of the phase angle $\xi \in [0,2\pi)$ is divided into 105 equally spaced points. 
For each value of $\xi$, we start with a random spin configuration at high temperature, $T = 7$ and decrease to $T = 2$ in temperature decrements of size $\delta T = 0.1$, followed by a slower cooling down in steps of size $\delta T = 0.01$ to a base temperature of $T = 0.01$. 
In this way of cooling the system, at each temperature, we perform $5\times 10^4$ Monte Carlo sweeps for equilibration and then, measure the thermal averages of the above six order parameters as well as the specific heat, $C_v$~\footnote{Specific heat at temperature $T$ from Monte Carlo simulations is computed as $C_v = \frac{\langle E^2\rangle - \langle E \rangle^2}{T^2}$ where $\langle E \rangle$ and $\langle E^2 \rangle$ are Monte Carlo averages of energy and energy-square at $T$, respectively.}, over $10^6$ MC samples, skipping three MC sweeps in between consecutive measurements.

To estimate the error bars on $C_v$ and on the order parameters, the $10^6$ measurements are divided into 25 blocks and then resampled using the standard bootstrap method~\cite{Newman1999}. 
Roughly $O(10^3)$ bootstrap samples were generated from these blocks to estimate the statistical errors. 
We perform the simulations for $L = 16$ and find that as the temperature approaches zero, the thermal average of the order parameter defined for a particular regime goes to unity in that regime~\footnote{In regions where it is not the relevant order, the order parameter is finite but very small, tending to zero in the thermodynamic limit}. 
We present the specific heat vs temperature and thermal average of the (regime-specific) order parameter vs temperature data with three different $\xi$ values belonging to three different regimes in Fig.~\subref{fig.sp-heat-ord-params}{(a)}, \subref{fig.sp-heat-ord-params}{(b)}, and \subref{fig.sp-heat-ord-params}{(c)}. 
Thus, across the whole phase diagram, the MC simulations reproduce the phase diagram obtained by the low-temperature expansion described in Sec.~\ref{ssec.obtd} [see Fig.~\ref{fig.phase-diagram}]. See Table~\ref{tab.summary} for a summary of the low-temperature phases. 

We briefly mention here some of the key points known about the phase at the special parameters $\xi = 0, \pi/2, \pi$, and their Klein duals, i.e., the six phase boundaries. In the thermodynamic limit, there is no long-range ordering at any non-zero temperature with $\xi = 0, \pi$, or at their Klein duals, as the Mermin-Wagner-Hohenberg theorem forbids spontaneously broken $O(3)$ continuous symmetry at nonzero temperature in two dimensions
~\cite{Mermin1966, Hohenberg1967}.
These disordered phases would be eliminated by any infinitesimally weak anisotropy, giving way to long-range ordered phases. At the compass points, a directional or nematic ordering along the director $\vhat{x}$ or $\vhat{y}$ is found at low temperatures ($\cramped{T < T_\textrm{dir}}$), with a phase transition into a disordered paramagnetic phase at high temperatures ($\cramped{T>T_\textrm{dir}}$)~\cite{Mishra2004,Wenzel2008}~\footnote{This phase transition falls in the two-dimensional Ising model universality class~\cite{Mishra2004,Wenzel2008}.}. 
Ref.~[\onlinecite{Mishra2004}] argues that the nematic ordering reduces to a conventional long-range ordering at low temperatures in the presence of a weak $XY$ exchange which favors long-range order along the $\pm\vhat{x}$ or $\pm\vhat{y}$ directions. In the context of the present study, the relevant non-compass perturbation would be the Heisenberg interaction. However, the nematic ordering should occupy a finite temperature fan ($\cramped{T_\textrm{dir}> T> J}$) that extends away from the compass points.

It is of interest to consider the critical temperature, $T_c$, for the transition between the high-temperature paramagnetic phase and the low-temperature ordered phase in the six regimes found from the MC simulations. 
For each $\xi$, this temperature is obtained from the location of the peak of the specific heat data. 
We present $T_c$ vs $\xi$ data for two different system sizes, $L =16$ and 20, in Fig.~\subref{fig.sp-heat-ord-params}{(d)}. 
The existence of a single peak in $C_v$ in the wide range of temperature shown in Fig.~\subref{fig.sp-heat-ord-params}{(d)} and the smooth change of the order parameter with $T$ below the peak indicates the presence of a single phase transition from the paramagnetic phase to the ordered phase at $T_c$. 
Thus, for each $\xi$, there is a single magnetic ordered phase below $T_c$, starting to develop at $T_c$ and gradually strengthening as $T\rightarrow 0$, with no further phase transition at any temperature below $T_c$ to any other phases than the one that developed at $T_c$. In other words, there is no further phase transition at any temperature below $T_c$ to any other phases from that developed just below $T_c$. As mentioned earlier, this single magnetic ordered phase below $T_c$ is consistent with the ordering found from the low temperature expansion valid at $T\ll T_c$ described in Sec.~\ref{ssec.obtd}.

It may seem surprising at first sight that the numerically observed critical temperatures at $\xi = 0, \pi$, and at their Klein dual points are significantly different from zero [see Fig.~\subref{fig.sp-heat-ord-params}{(d)}], where there should \emph{not} be any finite temperature phase transition in accordance with the Mermin-Wagner-Hohenberg theorem~\cite{Mermin1966,Hohenberg1967}\footnote{Note that at $\xi = 0, \pi$ and at their Klein dual points, the model [Eq.~\eqref{eq.Ham}] is an $O(3)$ Heisenberg model, and \emph{not} an $XY$ model that would have a Berezinskii–Kosterlitz–Thouless-type transition}. 
The apparent non-zero $T_c$ at those points is a finite-size effect, which should go to zero in the thermodynamic limit $\cramped{(N\rightarrow\infty)}$. 
However, eliminating these finite size effects would be quite computationally challenging~\cite{Schmoll2021} due to the infrared fluctuations diverging only \emph{logarithmically} in $L$
Thus, showing the vanishing of the transition temperature at these isotropic points in the parameter space would require significantly larger system sizes than the ones considered here. 
For other $\xi$ values, the Hamiltonian is anisotropic and thus the Mermin-Wagner-Hohenberg theorem does not prohibit ordering at non-zero temperature~\footnote{Note that in Regime-I, II, IV, and V, potential infrared divergence from the low dimensionality of the lattice and accidental continuous ground state degeneracy gets eliminated by order by thermal disorder induced anisotropy, and the long-range ordering prevails.
See Ref.~[\onlinecite{Khatua2023}] for a discussion on the subtle competition between infrared-divergent fluctuations and the effective anisotropy induced via order by thermal disorder.}.
Therefore, $T_c$ is expected to converge as the system size takes a moderately large value away from these fine-tuned $\xi=0$, HAF*, $\pi$, and HF* isotropic limits. 
This can be seen in the simulation results; the almost overlapping $T_c$ data for $L = 16$ and 20 in Fig.~\subref{fig.sp-heat-ord-params}{(d)} suggests that $T_c$ has nearly converged even at the relatively small size of $L = 16$ for all $\xi$ values except for $\xi = 0, \pi$, and their Klein dual points where $T_c$ should vanish in the thermodynamic limit.

\section{Quantum ground state phase diagram}
We have described above the classical phase diagram of the Heisenberg-compass model using several complementary methods: Luttinger-Tisza analysis, low-temperature spin-wave expansion and Monte Carlo simulations. 
In this section, we explore the quantum ground state phase diagram of this model. To begin, we examine the role of quantum fluctuations on the classical ground states at zero temperature. We commit particular attention to the regions where the classical model displays an accidental degeneracy and quantum ObD is expected to determine the ordering pattern.

\subsection{Order by disorder from quantum fluctuations at $T = 0$}
\label{ssec.qobd}

We start by considering the effects of quantum fluctuations perturbatively in the spin-length, approaching from the classical limit $\cramped{S \rightarrow \infty}$. 
This can be done through a quantum linear spin-wave analysis which introduces quantum fluctuations atop the classical ground state. This allows one to examine how accidental classical ground-state degeneracies may be lifted by these quantum fluctuations. 
 As mentioned in Sec.~\ref{ssec.obtd}, we discuss in Appendix~\ref{app.SW-analysis} the formalism for a quantum linear spin-wave analysis for various regimes of the phase angle $\xi$. 
 State selection via ObD due to quantum fluctuations at zero temperature for a particular $\xi$ is determined by the zero-point energy of the linear spin waves about an accidentally degenerate ground state parametrized by angle $\phi$, $\epsilon_{\rm{Q}}(\phi) = (1/2)\sum_{\vec{q}}\omega_{\vec{q}}(\phi)$. As mentioned previously in Sec.~\ref{ssec.obtd}, $\omega_{\vec{q}}(\phi)$ implicitly depends on $\xi$. The accidental ground state for which the zero-point energy is minimized is selected by quantum fluctuations, resulting in quantum ObD. 
We find that in Regime-I, II, IV, and V, the zero-point energy distinguishes between different accidentally degenerate classical ground states and selects a long-range ordered pattern.
Interestingly, in each of the above four regimes, the zero-point energy is found to be minimized for the very \emph{same} set of states chosen by the classical thermal ObD mechanism at low temperatures. 
Thus, ObD from quantum fluctuations at zero temperature predicts the same phase diagram as was found from the classical low-temperature expansion.

\subsection{Order by disorder from combined quantum and thermal fluctuations at $T >0$}
\label{ssec.qobd-thermal}
We have found that quantum ObD at zero temperature, and classical ObD from thermal fluctuations at low temperatures select the same long-range magnetic orders. However, we have not yet investigated ObD from \emph{combined} quantum and thermal fluctuations at small non-zero temperatures. This regime would appear when $\cramped{T \sim O(\omega_{\vec{q}})}$; between the low-temperature quantum limit ($\cramped{T \ll \omega_{\vec{q}}}$) and the classical spin-wave limit ($\cramped{\omega_{\vec{q}} \ll T \ll T_c}$) as discussed in Section~\ref{ssec.obtd}~\footnote{In typical spin systems where the spin-length lies between $S=1/2$ and $S = 7/2$, not all of these regimes will be practically realizable.}. 
ObD state selection at zero temperature and non-zero temperatures at $O(1/S)$ do not necessarily need to be the same, as was found in Refs.~\cite{Schick2020,PhysRevB.107.214414,hickey2024orderbydisorderquantumzeropointfluctuations}. Therefore, it is important to explore ObD state selection at $\cramped{T>0}$ including \emph{both} quantum and thermal fluctuations. 
For this purpose, we focus on the free energy of the quantum linear spin waves, which is given by~\cite{Kardar2007} 
\begin{eqnarray}
F_{\rm{Q}}(\phi) = \frac{1}{2}\sum_{\vec{q}}\omega_{\vec{q}}(\phi)+ T\sum_{\vec{q}}\ln\left(1 - e^{-\omega_{\vec{q}}(\phi)/T}\right),
\label{eq.free_energy}
\end{eqnarray}
with $\omega_{\vec{q}}(\phi)$ being the known spin-wave spectrum about an accidentally degenerate ground state characterized by $\phi$ for a given $\xi$. The first and second terms in Eq.~\eqref{eq.free_energy} correspond to the zero-point energy, $\epsilon_{\rm Q}(\phi)$, and the non-zero temperature contributions of the quantum spin waves to the free energy, respectively. ObD selects the states for which the free energy [Eq.~\eqref{eq.free_energy}] is minimized. We numerically compute the free energy as a function of $\phi$ for any $\xi$ in Regime-I and Regime-IV, and find that it is minimized for the same states as those selected by the zero-point energy alone, described in Sec.~\ref{ssec.qobd}. That is, the N\'eel states along $\pm\vhat{x}, \pm\vhat{y}$ for any coupling parameter in Regime-I, and the ferromagnetic states along $\pm\vhat{x}, \pm\vhat{y}$ for any coupling parameter in Regime-IV get selected by ObD at non-zero temperatures. 
By using the Klein duality, we conclude thatStripe-$\parallel$ and Stripe-$\perp$ are selected by ObD in Regime-II and V, respectively.    

In summary, quantum ObD at zero temperature, quantum-thermal ObD at non-zero temperature, and classical thermal ObD all select the same states. In other words, as has been found in many systems exhibiting ObD~\cite{Henley1989,Chaloupka2010,Price2012,Savary2012,McClarty2014}, but is not guaranteed~\cite{Schick2020,PhysRevB.107.214414,hickey2024orderbydisorderquantumzeropointfluctuations}, quantum and thermal fluctuations do \emph{not} compete in their respective selection of ground states via ObD.  

\begin{figure}
\includegraphics[width=\columnwidth]{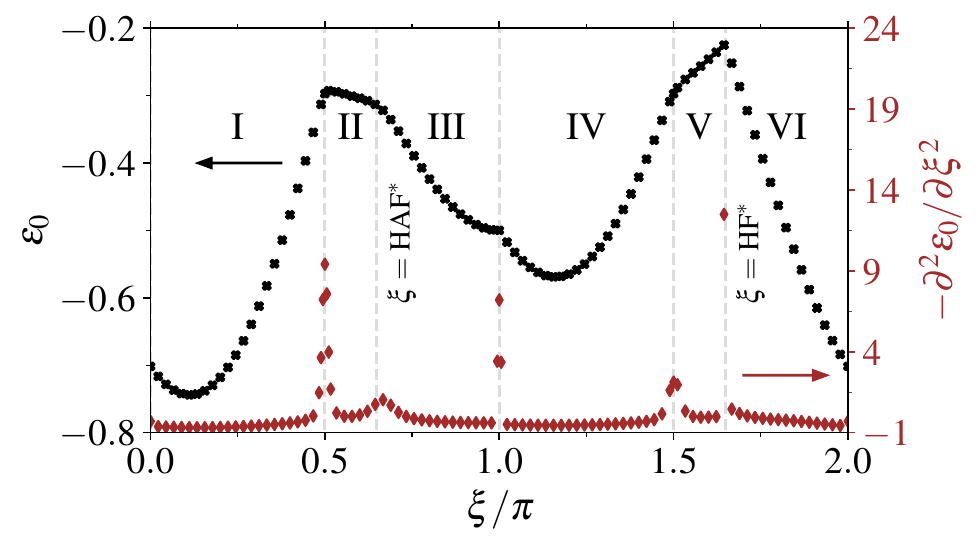}
\caption{Ground state energy per spin, $\varepsilon_0\equiv E_0/N$ where $E_0$ is the ground state energy eigenvalue, and the second derivative of $\varepsilon_0$ with respect to the phase angle $\xi$ for $L = 4$. The second derivative is computed numerically using a finite-difference formula using $\varepsilon_0$ obtained from exact diagonalization. The spikes in the second derivative provide the quantum phase boundaries, exhibiting excellent agreement with the classical phase boundaries (grey dashed lines). Regime-I to VI are labelled in accordance with Fig.~\ref{fig.phase-diagram}.}
\label{fig.e0-grad2}
\end{figure}

\begin{figure*}
\includegraphics[width=\textwidth]{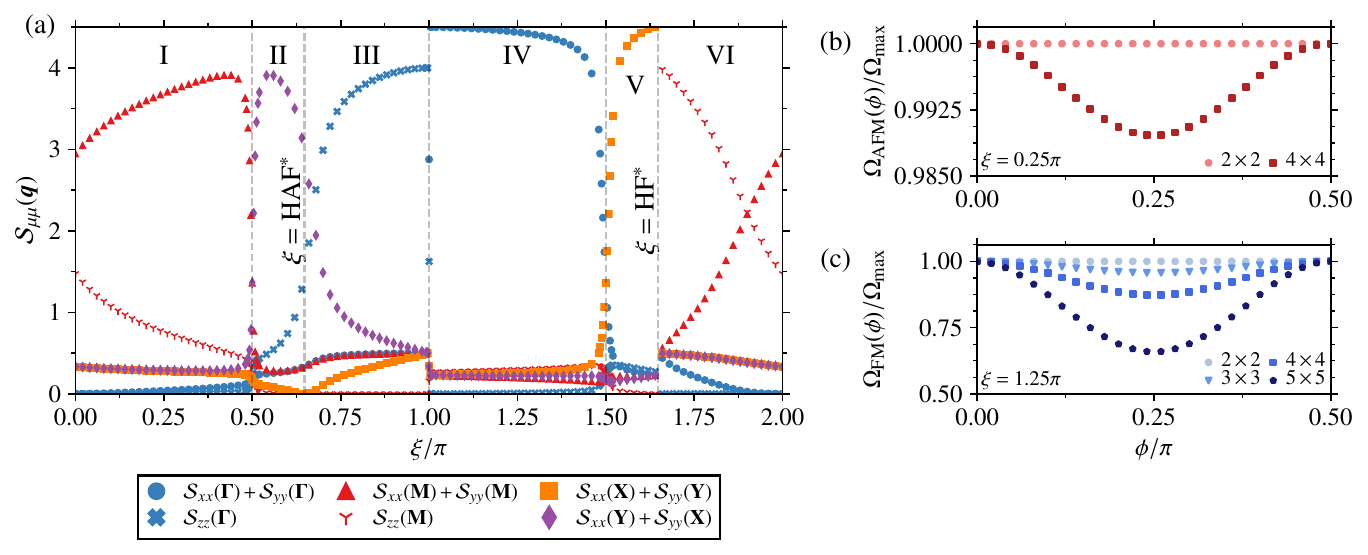}
\caption{(a) Results obtained from exact diagonalization of the quantum Heisenberg-compass model [Eq.~\eqref{eq.Ham}] on the $4\times4$ square lattice with periodic boundary conditions. Several static structure factors [Eq.~\eqref{eq.ssf}] computed over the full range of $\xi$ at the four high symmetry points of the Brillouin zone, $\mathbf{\Gamma} = (0,0),\mathbf{X} =(\pi,0),\mathbf{Y} = (0,\pi)$, and $\mathbf{M} = (\pi,\pi)$. Well-defined crossovers in the dominant structure factor are in excellent agreement with the classical phase boundaries (grey dashed lines).  Regime-I to VI are labelled in accordance with Fig.~\ref{fig.phase-diagram}. (b) Ratio of the ground state overlap with the Néel coherent state along $\phi$, $\Omega_{\rm AFM}(\phi)$, to the maximum of the overlap over the full range of $\phi$, $\Omega_{\rm max}$, with $\xi=0.25\pi$ for several system sizes in Regime-I. (c) Ratio of the ground state overlap with the ferromagnetic coherent state along $\phi$, $\Omega_{\rm{FM}}(\phi)$, to the maximum of the overlap over the full range of $\phi$, $\Omega_{\rm max}$, with $\xi=1.25\pi$ for several system sizes in Regime-IV. The range of $\phi$ is restricted to be only $[0,\pi/2)$ due to the $C_4$ symmetry of the model.}
\label{fig.ssf-plt}
\end{figure*}

\subsection{Numerical exact diagonalization}
 Given the material examples of interest have spin $S=1/2$, the semi-classical results derived at large-$S$ must be corroborated by direct calculations in the small-$S$ limit.
To this end, we investigate the quantum spin-$1/2$ Heisenberg-compass model [Eq.~\eqref{eq.Ham}] using exact diagonalization~\cite{Sandvik2010} on small clusters with periodic boundary conditions. 
For a system of $N = L^2$ spins, the dimension of the Hilbert space is given by $2^N$, which grows rapidly with system size and limits the size of cluster that can be realistically considered. 
To consider clusters that are as large as possible, we exploit the discrete translation symmetry of the model which divides the Hilbert space into $L^2$ momentum sectors which block diagonalize the Hamiltonian (see Ref.~\cite{Sandvik2010} for details).
These blocks can be diagonalized individually using sparse diagonalization methods~\cite{Wu2000}, such as the Lanczos algorithm, which can efficiently extract the low-lying energy eigenvalues and eigenvectors. 
Using this approach we are in principle able to consider system sizes up to $L=5$ ($N=25$) with reasonable computational effort.
However, since the ground state orderings that we expect in Regime-I, II, V, and VI have two sublattice magnetic unit cells, these are only compatible with even values of $L$, limiting a global phase diagram to only $L=4$.
However, the ferromagnetic orderings \emph{are} compatible with the $L = 5$ case and thus Regimes III and IV can be studied at this larger size.
To capture the full phase diagram, we primarily present the results obtained using exact diagonalization performed on a square lattice of size $L =4$, which is compatible with the magnetic ordering in all parameter regimes, unless otherwise specified.

We first determine the ground state and its energy, $E_0$, for a range of $\xi$ values across the full parameter space for $L =4$.  In Fig.~\ref{fig.e0-grad2}, we illustrate how the ground state energy per spin, $\varepsilon_0 = {E_0}/{N}$ varies with $\xi$. 
Near a quantum phase transition, we expect the ground state energy to change sharply with $\xi$~\cite{Sachdev2011} and thus to identify changes in the ground state phase, we consider not just $\varepsilon_0$ itself, but also its derivatives with respect to $\xi$.
Explicitly, to clearly identify the points at which $\varepsilon_0$ is changing quickly, we numerically computed the second derivative, $-{\partial^2 \varepsilon_0}/{\partial \xi^2}$ using a finite-difference formula.
This quantity is expected to show sharp peaks near any phase transition; for example, for a level crossing (first order transition) we would expect a discontinuity in the first derivative of $\varepsilon_0$ and thus a $\delta$-function in the second derivative. 
Consequently, the peaks in $-{\partial^2 \varepsilon_0}/{\partial \xi^2}$ [shown in Fig.~\ref{fig.e0-grad2}] provide good indicators for the locations of the boundaries between different phases. 
As depicted in Fig.~\ref{fig.e0-grad2}, $\varepsilon_0$ exhibits pronounced kinks near $\xi = \pi/2, \pi$, and HF*. 
Looking at the second derivative, we find that there are pronounced peaks at $\xi = \pi/2, \pi$ and HF*, suggesting transitions at those points, and less pronounced peaks at $\xi = 0,$ HAF*, and $3\pi/2$, suggesting weaker transitions~\cite{Chaloupka2010}.
Note that the peak in the second derivative between Regime-II and Regime-III is not exactly at HAF*, but slightly shifted to the right. 
However, Klein duality dictates that if HAF is a phase boundary, so is HAF*. 
That HAF is a phase boundary can be understood in both the quantum and classical cases through the breaking of the SU(2) spin rotation symmetry –- picking either an XY (in Regime-I) or Ising-like (in Regime-VI) ordered phase near the HAF point.
Therefore, the peak in $-{\partial^2 \varepsilon_0}/{\partial \xi^2}$ in Fig.~\ref{fig.e0-grad2} between Regime-II and Regime-III is expected to be exactly at HAF*. 
On the basis of the argument above, we believe the discrepancy between the peak position and HAF* is not due to any (peculiar) quantum effects but, rather, it is likely due to finite size effects, which should diminish as the system size increases. 
Apart from this slight discrepancy, the locations of the peaks are in excellent agreement with the phase boundaries obtained from the classical or semi-classical methods, and what one expects based on the constraints from the Klein duality. 
Therefore, given the system sizes for which we are able to perform exact diagonalization, the same number of phases are observed in the quantum spin-1/2 version of the Heisenberg-compass model as seen in its classical counterpart.  

With the phase boundaries identified, we next investigate the nature of the quantum ground state obtained via exact diagonalization. 
To this end, we consider the spin-spin correlations within the ground state, particularly, those encoded in the diagonal elements of the static structure factor,
\begin{equation}
\label{eq.ssf}
\mathcal{S}_{\mu\mu}(\boldsymbol{q})=\frac{1}{N}\sum_{\boldsymbol{r},\,\boldsymbol{r}'}e^{-i\boldsymbol{q}\cdot(\boldsymbol{r}-\boldsymbol{r}')}\langle S_{\boldsymbol{r}\phantom{'}}^{\mu}
S_{\boldsymbol{r}'}^{\mu} \rangle,
\end{equation}
where $\mu = x,y,z$ and $\avg{\cdots}$ is the expectation value in the ground state.
The structure factors are computed at several wave vectors in the first Brillouin zone, including four high symmetry points, $\mathbf{\Gamma} = (0,0),\,\mathbf{X} =(\pi,0),\,\mathbf{Y} = (0,\pi)$, and $\mathbf{M} = (\pi,\pi)$. We find that apart from these four special points, the structure factors at all other wave vectors are not very intense. We thus present the results only for the wave vectors: $\mathbf{\Gamma},\mathbf{X},\mathbf{Y},$ and $\mathbf{M}$ in Fig.~\subref{fig.ssf-plt}{(a)}. 
The combinations, $\mathcal{S}_{xx}(\mathbf{M})+\mathcal{S}_{yy}(\mathbf{M})$ and $\mathcal{S}_{xx}(\mathbf{\Gamma})+\mathcal{S}_{yy}(\mathbf{\Gamma})$ are the largest in Regime-I and IV, respectively. This suggests that the ground state is largely antiferromagnetically and ferromagnetically ordered in the $\vhat{x}-\vhat{y}$ plane in Regime-I and IV, respectively. 
Similarly, in Regime-II and V, $\mathcal{S}_{xx}(\mathbf{Y})+\mathcal{S}_{yy}(\mathbf{X})$ and $\mathcal{S}_{xx}(\mathbf{X})+\mathcal{S}_{yy}(\mathbf{Y})$ are the largest, respectively. 
This indicates that the orderings in Regime-II and V are well described by Eq.~\eqref{eq.case2} and Eq.~\eqref{eq.case5}, respectively. 
In Regime-III and VI, $\mathcal{S}_{zz}(\mathbf{\Gamma})$ and $\mathcal{S}_{zz}(\mathbf{M})$ are the largest, respectively. The orderings are thus largely ferromagnetic and antiferromagnetic along the $\pm\vhat{z}$ directions in Regime-III and VI, respectively. 
In summary, the structure factors reveal that the quantum ground states exhibit ordering wave vectors and (staggered) magnetization directions consistent with the classical ground states discussed in Sec.~\ref{sec.cgs}. 

The nature of the phases in the spin-1/2 model near the compass points ($\xi = \pi/2$ and $3\pi/2$) deserves some discussion. At the compass points, an exact diagonalization study~\cite{Dorier2005} has found a set of  $2^{(L+1)}$ low lying states that collapse into degenerate ground states as the system size is increased. A quantum Monte Carlo study~\cite{Wenzel2008} reveals that the spin-1/2 case also exhibits a directional or nematic ordering transition at \emph{finite} temperature, similarly to the classical result. We are not aware of any detailed study of the robustness of the directional-ordered phase at non-zero temperature against weak symmetric perturbations such as the Heisenberg interaction considered here. How these perturbations affect the directional-ordered phase in the vicinity of the pure compass limit -- and in particular its stability -- is a question we leave for future work.

We now examine the ObD state selection within exact diagonalization. 
Since quantum ground states of finite systems do not exhibit spontaneous symmetry breaking, they can be more usefully thought of as superpositions of states with definite ordering directions. 
ObD preference for specific orderings would result in having more weight on the states corresponding to those orderings in the superposition. 
For instance, the N\'eel states along the $\pm\vhat{x}, \pm\vhat{y}$ directions would be expected to have more weight than any other in-plane N\'eel state in the quantum ground state for any $\xi$ in Regime-I. 
To confirm this, we compute the overlap of the in-plane N\'eel states and the ground state wavefunction for a given $\xi$ in Regime-I obtained from exact diagonalization, $\ket{\Omega}$. 
A simple N\'eel state characterized by the in-plane angle $\phi$ is given by the product coherent state~\cite{Auerbach1998}, 
$$
\ket{\phi}_{\rm AFM} = \bigotimes_{\vec{r}}
\frac{1}{\sqrt{2}} \left(\ket{\uparrow} + (-1)^{\vec{r}} e^{i\phi} \ket{\downarrow}\right)
$$ 
where $\ket{\uparrow}$, $\ket{\downarrow}$ are $\vhat{z}$ quantized spin-1/2 states and $\vec{r}$ labels the sites of the lattice.
The overlap of this N\'eel state with $\ket{\Omega}$, that we refer to as $\Omega_{\rm AFM}(\phi) \equiv |\braket{\Omega|\phi}_{\rm AFM}|^2$, for $\xi = \pi/4$ is presented in Fig.~\subref{fig.ssf-plt}{(b)}, exhibiting maximal overlap with the N\'eel states corresponding to $\phi = 0, \pi/2, \pi, 3\pi/2$ (i.e., the N\'eel states along $\pm\vhat{x}, \pm\vhat{y}$). 
We also find that the anisotropy of this overlap increases as the system size increases. 
We note that while $\Omega_{\rm AFM}(\phi)$ does not explicitly represent the probability of finding the N\'eel coherent state along $\phi$ in the ground state due to the non-orthogonality of different coherent states, this quantity gives us a qualitative idea of how close the ground state is to a given product state. 
We find this to be true for any value of $\xi$ in Regime-I. 
This thus confirms the preference for ObD-selected Néel states along $\pm\vhat{x}, \pm\vhat{y}$ within the quantum ground state in this regime. Note the smallness of the change in the overlaps as a function of angle $\phi$ likely originates from the weakness of the ObD selection (discussed in Sec.~\ref{ssec.qobd}), which is several orders of magnitude smaller than the scale of $J$ or $K$.

We can proceed similarly in Regime-IV by considering overlaps~\footnote{In Regime-IV, we consider systems with an odd number of spins as well, in which the ground states are doubly degenerate, say $\ket{\Omega_1}$ and $\ket{\Omega_2}$, due to Kramers' theorem. Any linear combination of those two ground states is also a ground state. However, to characterize the relevant classical state, we consider a particular linear combination $
\ket{\Omega}=\cos\alpha\,\ket{\Omega_1}+\sin\alpha\,e^{i\beta}\,\ket{\Omega_2}$ which maximizes the overlap with a given $\ket{\phi}_{\rm{FM}}$.} with the in-plane ferromagnetic coherent states 
$$
\ket{\phi}_{\rm{FM}} =\bigotimes_{\vec{r}}
\frac{1}{\sqrt{2}} \left(\ket{\uparrow} +e^{i\phi} \ket{\downarrow}\right),
$$ 
defining $\Omega_{\rm FM}(\phi) \equiv |\braket{\Omega|\phi}_{\rm FM}|^2$ where $\ket{\Omega}$ now represents the ground state wavefunction for a given $\xi$ in Regime-IV obtained from exact diagonalization.
We show the results for various system sizes with $\xi=1.25\pi$ in Fig.~\subref{fig.ssf-plt}{(c)}. 
The overlap is found to be maximal for the in-plane angles $\phi=0,\pi/2,\pi,3\pi/2$, confirming the preference for the ObD selected ferromagnetic states along $\pm\vhat{x}, \pm\vhat{y}$ within the quantum ground state. 
While Fig.~\subref{fig.ssf-plt}{(c)} is shown for a particular $\xi$ [i.e., $\xi=1.25\pi$], the same was found to be true for any $\xi$ in Regime-IV. 
Using the Klein duality, we argue that the ObD-selected product states have dominant contributions to the quantum ground state wavefunctions in Regime-II and V, as well. 
Therefore, the quantum ground state phase diagram found from exact diagonalization on small systems like $4\times4$ square lattice is qualitatively similar to the low-temperature classical phase diagram and zero temperature semi-classical phase diagram discussed previously.

\section{Summary and Discussion}

In this work, we examined the zero-temperature quantum ground state phase diagram and the low-temperature classical and  quantum phase diagram of the Heisenberg-compass model on the square lattice. Notably, this model admits a  Klein duality which facilitates a mapping of the spin-spin interaction parameters from one set to another. This duality analysis partitions the entire parameter space into six distinct regimes, with three of them being Klein-dual counterparts of the remaining three. As a result, the properties of the model in a parameter regime are related to those in its dual regime. For two of those six regimes, the classical zero temperature ground states consist of two symmetry-related discrete configurations. In the remaining four regimes, the classical ground states display an accidental continuous degeneracy characterized by an $O(2)$ manifold. Using classical Monte Carlo simulations and spin-wave analysis, we analyzed the low-temperature classical phase diagram of this model. 
These calculations reveal six different ordered phases in the six parameter regimes, with four order by disorder (ObD) phases stemming from thermal fluctuations and two energetically ordered phases. By considering quantum fluctuations via a quantum spin-wave analysis at zero temperature, we find that ObD from quantum fluctuations stabilizes the same ordered states as those derived from the preceding classical methodologies. Furthermore, a calculation of the free energy from quantum spin waves at $T>0$ finds that the combined effect of thermal and quantum fluctuations at low temperatures favors the same states as those selected by the quantum fluctuations alone at zero temperature. Additionally, by investigating the zero-temperature quantum ground state phase diagram using numerical exact diagonalization on small finite clusters, we find identical phase diagram to the one obtained from the classical analysis and the quantum spin wave analysis.  

\subsection{Perspective on applications to materials} 
It is of interest to briefly discuss the relevance of our paper to real materials.
Generically, a magnetic material with its magnetic ions on a square lattice will not precisely correspond to the Heisenberg-compass model, even at the nearest-neighbor level, as it may possess additional symmetry-allowed interactions. Explicitly, the space group symmetries of the square lattice allow for an additional bond-independent Ising interaction, $S^z_{\vec{r}}S^z_{\vec{r}'}$, on each nearest-neighbor bond~\cite{Bertinshaw2019}. Incorporating such a term in the original Heisenberg-compass model of Eq.~\eqref{eq.Ham}, yields
\begin{equation}
   \label{eq.Ham-enlarged}
  {\cal H} = \sum_{\vec{r},\vec{\delta}}
  \biggl[J \vec{S}^{\phantom{x}}_{\vec{r}} 
  \!\cdot   \vec{S}^{\phantom{x}}_{\vec{r}  +\vec{\delta}\phantom{+\vec{}}}
    \!\! + K S^\delta_{\vec{r}\phantom{+\vec{}}} 
    \! \! S^\delta_{\vec{r}+\vec{\delta}}
    +\Delta\, S^z_{\vec{r}\phantom{+\vec{}}} 
    \! \! 
    S^z_{\vec{r}+\vec{\delta}}
    \biggr] ,
\end{equation}
where $\Delta$ parametrizes the strength of $S^z_{\vec{r}}S^z_{\vec{r}'}$ Ising anisotropy. Interestingly, we note that even in the presence of the additional $S^z_{\vec{r}}S^z_{\vec{r}'}$ interaction, a version of the Klein duality still holds, with the rotation described in Sec.~\ref{sec.klein-duality} providing an exact mapping between two parameter sets: $(J, K, \Delta) \rightarrow (-J, 2J+K, -\Delta)$.

Importantly, a small $\Delta$ ($\cramped{\Delta\ll J}$ and $\cramped{\Delta\ll K}$), either positive or negative, \emph{does not} lift any of the aforementioned in-plane accidental classical $O(2)$ degeneracies, and therefore does not qualitatively affect the order by disorder physics of the Heisenberg-compass model.
This is a generic statement for \emph{any} symmetry allowed bilinear spin-exchanges, as is found in other order-by-disorder material candidates, such as Er$_2$Ti$_2$O$_7$~\cite{Savary2012}. Adapting the arguments of Ref.~[\onlinecite{Savary2012}], the classical energy of any of the in-plane, accidentally degenerate states is characterized by an order parameter $\vec{m} \equiv (m_x,m_y)$ (listed in Table~\ref{tab.summary}) which transforms as $(m_x,m_y) \rightarrow (m_y, -m_x)$ under the $C_4$ symmetry. 
It is straightforward to show that the only bilinear energy function that can be constructed from $\vec{m}$ is $\propto |\vec{m}|^2$ which enjoys an accidental $O(2)$ symmetry. We thus see that these accidental degeneracies, and thus the order-by-disorder, persists even in the presence of generic symmetry allowed bilinear interactions of arbitrary range and should thus be relevant in realistic material-relevant extensions of the Heisenberg-compass model.

As reported in Ref.~\cite{Katukuri2014}, a single layer of the perovskite iridate Ba$_2$IrO$_4$ can effectively be described by the Heisenberg-compass model on the square lattice. 
Using \emph{ab initio} quantum-chemistry computational techniques, the authors of Ref.~[\onlinecite{Katukuri2014}] estimated exchange couplings $J$ $\approx$ 65 meV and $K$ $\approx$ 3.5 meV, positioning the system within Regime-I of the current study. 
Notably, the authors of Ref.~[\onlinecite{Katukuri2014}] found a negligible value for the Ising anisotropy $\Delta$ in Eq.~\eqref{eq.Ham-enlarged}. 
The smallness of this coupling could potentially be attributed to the enhanced symmetries that appear when one restricts the \emph{ab initio} calculations to exchange paths considering two ideal neighboring IrO$_6$ octahedra. 
Unlike the full layer, this pair of octahedra possesses an additional $C_4$ symmetry about the bond axis which, if exact, forbids any Ising anisotropy (but allows for a compass interaction, $K$). Therefore, this $\cramped{\Delta\, S^z_{\vec{r}\phantom{+\vec{\delta}}}\!\!\!\!S^z_{\vec{r}+\vec{\delta}}}$ coupling should only be generated by exchange processes that go beyond the pair octahedra or involve tetragonal distortions of the IrO$_6$ octahedra in the out-of-plane direction.
The limit of small or perturbative $\Delta$, where our results are valid, is thus relevant in the context of Ba$_2$IrO$_4$~\footnote{Note that in materials such as Ba$_2$IrO$_4$, the two-dimensional square lattices are $AB$ stacked in such a way that the four-fold rotation permutes the sublattices on alternating layers~\cite{Katukuri2014}. This complicates the Landau-type analysis of the effect of symmetry-allowed inter-layer exchanges discussed in the previous paragraph.}.

Although our results are relevant for understanding the ground-state properties of a single-layer Ba$_2$IrO$_4$ [$J$ $\approx$ 65 meV, $K$ $\approx$ 3.5 meV, and perturbative $\Delta$]~\cite{Katukuri2014}, they do not explain the experimentally observed magnetic ordering in Ba$_2$IrO$_4$ -- N\'eel order along [110] direction~\cite{Boseggia2013} (as opposed to the expected ordering in Regime-I of our study). To explain the experimentally observed ordering, Ref.~[\onlinecite{Katukuri2014}] argues that, along with nearest-neighbour exchanges, one needs to take into account a subset of the inter-layer exchanges which competes with the intra-layer exchanges. However, it has been argued that in Sr$_2$IrO$_4$~\cite{Liu2019, Kim2022} the combined effects of strong spin-orbit coupling and Jahn-Teller distortion acting within a layer are responsible for the experimentally observed ordered moment direction. In light of these results and the similar material context, it would be interesting to extend the work presented in this study to consider the competition between energetic and ObD state selection in a model of Ba$_2$IrO$_4$ incorporating both the symmetry-allowed in-plane $(J,K,\Delta)$ interactions as well as spin-lattice and other couplings discussed in more detail for Sr$_2$IrO$_4$~\cite{Liu2019,Kim2022}.

\subsection{Avenues for future work on the Heisenberg-compass model} 
Since the Heisenberg-compass model displays four distinct parameter regimes exhibiting both thermal and quantum ObD, it provides a rich playground for investigating the conceptual underpinnings of ObD more broadly. Within this model, the following open avenues may be of particular interest to explore further:
\begin{enumerate}[label={(\alph*)},leftmargin=*] 
\itemsep0em
    \item  ObD induces a dynamically generated pseudo-Goldstone gap~\cite{Weinberg1972,Burgess2000} in the excitation spectrum. This gap has been computed previously in the context of ObD from quantum fluctuations at $T = 0$~\cite{Rau2018} and from purely thermal fluctuations at non-zero $T$~\cite{Khatua2023} in order to explore and expose \emph{model-independent universal} signatures of ObD. However, the characteristics of the pseudo-Goldstone gap arising from ObD due to \emph{combined} thermal and quantum fluctuations at $T>0$,  perhaps the most general and relevant scenario for real magnetic systems, have, to the best of our knowledge, not yet been systematically explored in the literature. The Heisenberg-compass model offers an opportunity to investigate this in all four of its ObD regimes.
    \item  A magnet with long-range order may exhibit excitations different from conventional magnons, such as two-magnon bound states~\cite{Bethe1931, Wortis1963, Keselman2020,Kato2020, Mook2023}. Such quasi-particle excitations may, for example, impact the heat transport at low energies~\cite{Subrahmanyam2004}. Study of such bound states in magnets with long-range order arising from ObD has, again to the best of our knowledge, remained unexplored. Since the pseudo-Goldstone gap generated by ObD may be typically small, there may exist two-magnon bound states of energy scale comparable to the gap, which could significantly impact the low energy properties of a system harboring ObD. 
    \item  Much of the literature on ObD has focused on the classical or semi-classical limit ($\cramped{S\rightarrow \infty}$) and does not readily apply in the more realistic quantum limit (e.g. $S=1/2$). A well-understood and unbiased method to study spin-1/2 systems is exact diagonalization~\cite{Sandvik2010}. As this method is limited to small systems, observables often exhibit large finite-size effects. Therefore, one avenue to better understand ObD and its dynamical implications in spin-1/2 systems may be to characterize the finite-size manifestations of ObD and understand how they might appear in exact diagonalization calculations. 
    The topic of finite-size signatures of ObD has been little explored~\cite{Lecheminant1995, Khatua2021}, warranting further investigation. More practically, achieving an understanding of such finite-size signatures could be directly relevant to finite size real quantum magnetic systems, such as small magnetic flakes or molecular magnets~\cite{Park2016} and trapped ion quantum simulators~\cite{kotibhaskar2023}.   
\end{enumerate}

To conclude, we believe that the Heisenberg-compass model in two dimensions is a simple and compelling model to explore and shed some light on the above interesting theoretical questions. We look forward for theoretical developments in these, and perhaps other directions. These would deepen our understanding of ObD and help uncover ways to unambiguously expose its manifestation in real physical systems.

\begin{acknowledgments}
We thank Itamar Aharony, 
Kristian Tyn Kai Chung, 
R. Ganesh,  
Felipe G\'{o}mez-Lozada, 
Alex Hickey, 
Andreas L\"auchli, 
Daniel Lozano-G\'{o}mez  
and Natalia Perkins 
for useful discussions.
We also thank Giniyat Khaliullin for useful comments on an early version of the manuscript.
We acknowledge the use of computational resources provided by Digital Research Alliance of Canada.  
This research was funded by the NSERC of Canada~(M.J.P.G, J.G.R) and the Canada Research Chair Program~(M.J.P.G, Tier I). G.C.H acknowledges funding from the NSERC of Canada through the USRA and CGS-M programs.
\end{acknowledgments}

\appendix
\section{Luttinger-Tisza method for determination of the classical ground states}
\label{app.LT}
In this appendix, we describe the Luttinger-Tisza method~\cite{Kimchi2014,Luttinger1946,Niggemann2020} used to determine the classical ground states of the model of Eq.~\eqref{eq.Ham}. We start by rewriting the Hamiltonian in the following way,
\begin{equation}
    {\cal H} = \frac{1}{2}\sum_{\vec{r},\,\vec{\gamma}}\trp{\vec{S}}_{\vec{r}}\mat{J}_{\vec{\gamma}}\vec{S}_{\vec{r}+\vec{\gamma}},
    \label{eq.Ham1}
\end{equation}
where $\vec{\gamma} = \pm\vec{x},\,\pm\vec{y}$ denotes all four nearest-neighbor bonds.
The prefactor $1/2$ comes from the double counting of each bond, and the interaction matrices are  
\begin{equation}
    \mat{J}_{\vec{x}} = \mat{J}_{-\vec{x}}= \left(\begin{array}{ccc}
        J+K &  0 &0\\
        0 & J & 0\\
        0& 0& J
    \end{array}\right),
   \,\, \mat{J}_{\vec{y}} = \mat{J}_{-\vec{y}}= \left(\begin{array}{ccc}
        J &  0 &0\\
        0 & J+K & 0\\
        0& 0& J
    \end{array}\right). 
    \label{eq.exchange-global}
\end{equation} 
Under Fourier transform, $\vec{S}_{\vec{r}} = \big(1/\sqrt{N}\big)\sum_{\vec{q}}\vec{S}_{\vec{q}} e^{i\vec{q}\cdot\vec{r}}$, Eq.~\eqref{eq.Ham1} becomes
\begin{equation}
    {\cal H} = \frac{1}{2}\sum_{\vec{q}}\trp{\vec{S}}_{-\vec{q}}\mat{J}_{\vec{q}}\vec{S}_{\vec{q}},
    \label{eq.Ham-FT}
\end{equation}
where the Fourier transformed interaction matrix, $\mat{J}_{\vec{q}} = \sum_{\vec{\gamma}}\mat{J}_{\vec{\gamma}} e^{i\vec{q}\cdot\vec{\gamma}} = 2(\cos q_x)\mat{J}_{\vec{x}} + 2(\cos q_y)\mat{J}_{\vec{y}}$ using the fact, $\mat{J}_{\vec{x}} = \mat{J}_{-\vec{x}}$ and $\mat{J}_{\vec{y}} = \mat{J}_{-\vec{y}}$.
\begin{figure*}
    \centering
    \includegraphics[width=0.9\textwidth]{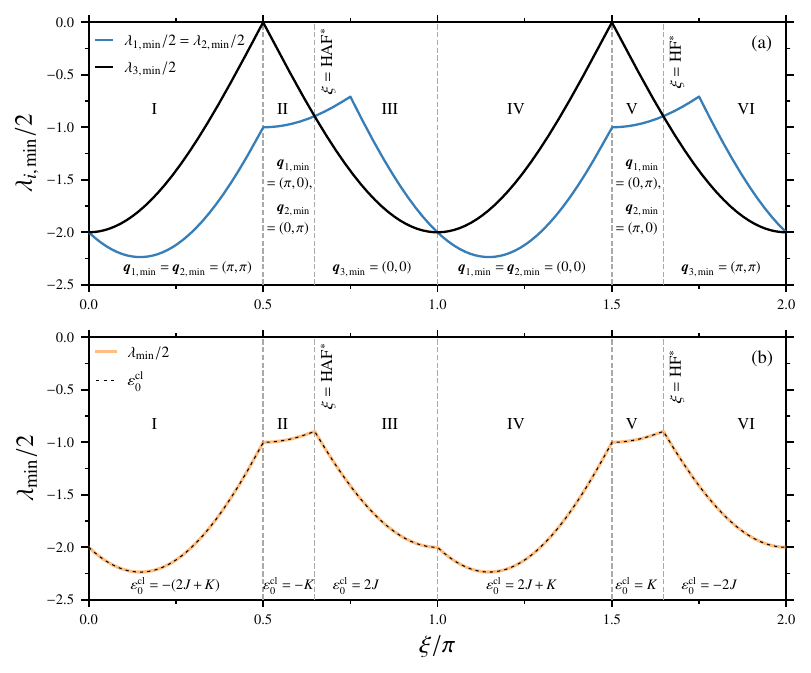}
    \caption{(a) Minima of the eigenvalues of the Fourier transformed interaction matrix over the first Brillouin zone, $\lambda_{i,\,{\rm min}}$ [where $i = 1,2,3$] multiplied by a factor of one-half shown for different $\xi$. By tracking the minimum of $\lambda_{1,{\rm min}}, \lambda_{2,{\rm min}},$ and $\lambda_{3,{\rm min}}$ at each $\xi$, $\lambda_{\rm min}$, the full $\xi$ range can be divided into six smooth regions separated by kinks in $\lambda_{\rm min}$, $(0,\pi/2), (\pi/2, \pi - \tan^{-1}(2)), (\pi - \tan^{-1}(2), \pi), (\pi, 3\pi/2), (3\pi/2, 2\pi - \tan^{-1}(2)),$ and $(2\pi - \tan^{-1}(2), 2\pi)$. Wave vectors of the corresponding $\lambda_{\rm min}$, $\vec{q}_{i,{\rm min}}$ are specified in each region. These six regimes are labelled as Regime-I to VI separated by grey dashed lines. (b) $\lambda_{\rm min}/2$ plotted against $\xi$. At each of the above six regimes (Regime-I to VI), the functional form of $\lambda_{\rm min}/2$, $\varepsilon_0^{\rm cl}(J,K)$ is specified, giving the classical ground state energy per site as a function of $J$ and $K$.}
    \label{fig:lambda-min}
\end{figure*}
Using Eq.~\eqref{eq.exchange-global}, $\mat{J}_{\vec{q}}$ takes the form, 
\begin{equation}
    \mat{J}_{\vec{q}} = \left(\begin{array}{ccc}
        \lambda_{1}(\vec{q}) &  0 & 0\\
        0 &    \lambda_{2}(\vec{q}) & 0\\
        0 &  0&  \lambda_{3}(\vec{q})\\
    \end{array}\right),
\end{equation}
where 
\begin{eqnarray}
    \lambda_1(\vec{q}) &=& 2 (J+K)\cos q_x + 2 J \cos q_y\nonumber\\
    &=& 2 (\cos\xi+\sin\xi)\cos q_x + 2 \cos\xi \cos q_y, \nonumber \\
    \lambda_2(\vec{q}) &=& 2 J \cos q_x + 2 (J+K)\cos q_y \nonumber\\
    &=& 2 \cos\xi \cos q_x + 2 (\cos\xi+\sin\xi)\cos q_y , \nonumber \\
     \lambda_3(\vec{q}) &=& 2 J(\cos q_x + \cos q_y) \nonumber\\
     & =& 2 \cos\xi(\cos q_x +\cos q_y).
     \label{eq.lambda}
    \end{eqnarray}
Since $\mat{J}_{\vec{q}}$ is diagonal, $\lambda_1(\vec{q}), \lambda_2(\vec{q}), $ and $ \lambda_3(\vec{q})$ are its eigenvalues, and the corresponding eigenvectors are simply the Cartesian directions, i.e., $\vhat{x}, \vhat{y},$ and $\vhat{z}$. 
With this, the Fourier transformed Hamiltonian in Eq.~\eqref{eq.Ham-FT} takes the following form,
\begin{equation}
    {\cal H} = \frac{1}{2}\sum_{\vec{q}}\left(\lambda_1(\vec{q})\, |S_{\vec{q}}^x|^2 + \lambda_2(\vec{q})\, |S_{\vec{q}}^y|^2 + \lambda_3(\vec{q})\, |S_{\vec{q}}^z|^2   \right).
    \label{eq.Ham-eigval}
\end{equation}
The modulus square in Eq.~\eqref{eq.Ham-eigval} comes from the fact that $\vec{S}_{-\vec{q}} = \vec{S}^*_{\vec{q}}$ (spin components are real valued in direct space).
For a given $\xi$, let $\lambda_{\rm min}$ be the smallest of the minimum values of $\lambda_1(\vec{q}), \, \lambda_2(\vec{q}),$ and $ \lambda_3(\vec{q})$ across the first Brillouin zone (BZ). Then, Eq.~\eqref{eq.Ham-eigval} can be written as 
\begin{eqnarray}
    {\cal H} &=& \frac{1}{2}\sum_{\vec{q}}\left((\lambda_1(\vec{q}) - \lambda_{\rm min })\, |S_{\vec{q}}^x|^2 + (\lambda_2(\vec{q}) - \lambda_{\rm min })\, |S_{\vec{q}}^y|^2\right . \nonumber\\
    &&\hspace{0.7 cm} +\left. (\lambda_3(\vec{q}) - \lambda_{\rm min })\, |S_{\vec{q}}^z|^2   \right) + \frac{\lambda_{\rm min}}{2}\sum_{\vec{q}} |\vec{S}_{\vec{q}}|^2.
    \label{eq.LT}
\end{eqnarray}
Since $(\lambda_1(\vec{q}) - \lambda_{\rm min })$, $(\lambda_2(\vec{q}) - \lambda_{\rm min })$, and $(\lambda_3(\vec{q}) - \lambda_{\rm min })$ are all positive semi-definite, the classical ground state energy corresponds to the last term in Eq.~\eqref{eq.LT}, and is given by, 
\begin{equation}
E_{\rm min} = \frac{\lambda_{\rm min}}{2}\sum_{\vec{q}}|\vec{S}_{\vec{q}}|^2 = \frac{\lambda_{\rm min}}{2}\sum_{\vec{r}}|\vec{S}_{\vec{r}}|^2 = \frac{\lambda_{\rm min}}{2} N,
\label{eq.lt-emin}
\end{equation}
where $N$ is the total number of spins.
Note that, to finally obtain the classical minimum energy, $E_{\rm min}$, the Luttinger-Tisza method has made use of the ``weak'' spin-length constraint $\sum_{\vec{r}}|\vec{S}_{\vec{r}}|^2 = N$, and not the ``hard'' spin-length constraint $|\vec{S}_{\vec{r}}|^2 = 1$ for all $\vec{r}$.
Thus, we find that the ground state energy per site is $\lambda_{\rm min}/2$. 
The minima of $\lambda_1(\vec{q}), \, \lambda_2(\vec{q})$, and $\lambda_3(\vec{q})$ across the first BZ, denoted respectively as $\lambda_{1,{\rm min}}, \lambda_{2,{\rm min}},$ and $\lambda_{3,{\rm min}}$, are shown in Fig.~\subref{fig:lambda-min}{(a)} for the entire range of $\xi$. 
By tracking $\lambda_{\rm min}$ (i.e., minimum of $\lambda_{1,{\rm min}}, \lambda_{2,{\rm min}},$ and $\lambda_{3,{\rm min}}$), one finds six regions in the coupling parameter space divided by kinks as shown in Fig.~\subref{fig:lambda-min}{(a)}. This partition of the parameter space into six regimes is consistent with the six regimes found from the Klein duality in Sec.~\ref{sec.klein-duality}.
$\lambda_{\rm min}/2$ and its functional form with respect to $(J,K)$ are shown in Fig.~\subref{fig:lambda-min}{(b)}. 
In the following subsections, we shall determine the classical ground states of the model [Eq.~\eqref{eq.Ham}] in Regime-I and III using the Luttinger-Tisza method and, from these states determined, we can find the classical ground states in other regimes using the Klein duality transformation [Eq.~\eqref{eq.klein}] and the sublattice spin-flip transformation [Eq.~\eqref{eq.sublattice-flip}].

\subsection{Regime-I : $\xi\in (0,\pi/2)$}
\label{ssec.regime-I}
In this regime, $\lambda_1(\vec{q})$ and $\lambda_2(\vec{q})$ possess simultaneous global minima, $\lambda_{\rm min}$, at $\vec{q}_{\rm min} = (\pi,\pi)$ [see Fig.~\subref{fig:lambda-min}{(a)}] and the classical ground state energy per site is, $\lambda_{\rm min}/2 = -(2J+K)$ [see Fig.~\subref{fig:lambda-min}{(b)}]. 
Therefore, from Eq.~\eqref{eq.LT}, the minimum energy spin configuration would be such that 
\begin{eqnarray}
&S^{z}_{\vec{q}}& = 0\,\,\,\,\,\, {\rm{for\,\,\, all}} \,\,\vec{q},\nonumber\\
&S^x_{\vec{q}\neq\vec{q}_{\rm min}}& = S^y_{\vec{q}\neq\vec{q}_{\rm min}} = 0. 
\label{eq.regime-I-cond1}
\end{eqnarray}
Thus, from Eq.~\eqref{eq.LT}, the ground state energy can be written as
\begin{equation}
    E_{\rm min} = \frac{\lambda_{\rm min}}{2}\left(|S^{x}_{\vec{q}_{\rm min}}|^2 + |S^{y}_{\vec{q}_{\rm min}}|^2\right).
\end{equation}
Equating this to the ground state energy $\frac{\lambda_{\rm min}N}{2}$ leads to 
\begin{equation}
|S^{x}_{\vec{q}_{\rm min}}|^2 + |S^{y}_{\vec{q}_{\rm min}}|^2 = N.
\label{eq.regime-I-cond2}
\end{equation}
Using $\vec{S}^*_{\vec{q}} = \vec{S}_{-\vec{q}}$, at $\vec{q}_{\rm min} = (\pi,\pi)$ we have the property $\vec{S}^*_{(\pi,\pi)} = \vec{S}_{(-\pi,-\pi)} = \vec{S}_{(\pi,\pi)}$, where the last step involves $(\pi,\pi)\equiv (-\pi,-\pi)$. 
This shows that $\vec{S}_{\vec{q}_{\rm min}}$ is real. 
Using the real valuedness of $\vec{S}_{\vec{q}_{\rm min}}$ and using Eq.~\eqref{eq.regime-I-cond1} and Eq.~\eqref{eq.regime-I-cond2}, we can write $\vec{S}_{\vec{q}_{\rm min} = (\pi,\pi)} ~=~ \sqrt{N}(\cos\phi,\sin\phi,0)$ where $\phi\in[0,2\pi)$.
We can now obtain a description of the spins $\vec{S}_{\vec{r}}$ in the direct space by taking the inverse Fourier transform, 
\begin{align}
\vec{S}_{\vec{r}} = \frac{1}{\sqrt{N}}\sum_{\vec{q}}\vec{S}_{\vec{q}} e^{i\vec{q}\cdot\vec{r}} &= \frac{1}{\sqrt{N}}\vec{S}_{\vec{q}_{\rm min}=(\pi,\pi)} e^{i(\pi,\pi)\cdot\vec{r}}\nonumber \\
&= (-1)^{\vec{r}}(\cos\phi,\sin\phi,0).
\end{align}
We finally see that $\vec{S}_{\vec{r}} = (-1)^{\vec{r}}(\cos\phi,\sin\phi,0)$,
a N\'eel state in the $\vhat{x}-\vhat{y}$ plane with the N\'eel direction specified by the in-plane angle $\phi$. Note that these states satisfy the hard spin-length constraint, $|\vec{S}_{\vec{r}}|^2 = 1$, and are thus legitimate ground states produced by the Luttinger-Tisza method.

\subsection{Regime-III :  $\Big(\pi-\tan^{-1}(2)<\xi<\pi\Big)$}
\label{ssec.regime-III}
In this regime, $\lambda_3(\vec{q})$ has the global minimum, $\lambda_{\rm min}$, at $\vec{q}_{\rm min} = (0,0)$ [see Fig.~\subref{fig:lambda-min}{(a)}] and the classical ground state energy per site is, $\lambda_{\rm min}/2 = 2J$ [see Fig.~\subref{fig:lambda-min}{(b)}].
Therefore, from Eq.~\eqref{eq.LT}, the minimum energy configuration must satisfy $S^{x/y}_{\vec{q}} = 0$ for all $\vec{q}$ and $S^z_{\vec{q}\neq\vec{q}_{\rm min}} = 0$.
We can thus write the ground state energy from Eq.~\eqref{eq.LT} as 
\begin{equation}
E_{\rm min} = \frac{\lambda_{\rm min}}{2}|S^{z}_{\vec{q}_{\rm min}}|^2. 
\end{equation}
Equating this to the ground state energy $\frac{\lambda_{\rm min}N}{2}$ yields $|S^{z}_{\vec{q}_{\rm min}}|^2  = N$. 
The Fourier component of the real valued $\vec{S}_{\vec{r}}$ at $\vec{q}_{\rm min}=(0,0)$, $\vec{S}_{\vec{q}_{\rm min}}$, is real. 
All the above conditions on the Fourier transformed spins yields $\vec{S}_{\vec{q}_{\rm min}} = \sqrt{N}(0,0,\pm 1)$ which, in direct space, amounts to 
\begin{align}
\vec{S}_{\vec{r}} = \frac{1}{\sqrt{N}}\sum_{\vec{q}}\vec{S}_{\vec{q}} e^{i\vec{q}\cdot\vec{r}} = \frac{1}{\sqrt{N}}\vec{S}_{\vec{q}_{\rm min}=(0,0)} e^{i(0,0)\cdot\vec{r}} = (0,0,\pm 1).
\end{align}
Therefore, we find only two discrete states corresponding to ferromagnetic order along $\pm\vhat{z}$ directions. As we have $|\vec{S}_{\vec{r}}|^2 = 1$, these two states are legitimate ground states.

\section{Spin-wave analysis}
\label{app.SW-analysis}
In this appendix, we provide the details of the spin wave analysis of the Heisenberg-compass model [Eq.~\eqref{eq.Ham}] in the four parameter regimes [Regime-I, II, IV, V] that exhibit ObD, using the Holstein-Primakoff formalism~\cite{Auerbach1998}. We present the analysis only in Regime-I and IV, and then use the Klein duality to extend the results to the cases of Regime-II and V. 
We assume for the purpose of this analysis that there is only one sublattice in the magnetic unit cell in the classical ground state. 
While this is true for the ground state in Regime-IV (ferromagnetic state), the N\'eel ground state in Regime-I has two magnetic sublattices. 
However, we can make a transformation to the spins in Regime-I, changing the N\'eel state to a ferromagnetic state (one sublattice magnetic ordering) so that one sublattice spin-wave analysis can be applied to obtain the results in Regime-I as well.

We begin with the Hamiltonian written in a slightly different form as in Eq.~\eqref{eq.Ham1}. 
Assuming there is only one sublattice in the magnetic unit cell in the ground state, we define a local frame, $\left(\vhat{e}_x(\phi), \vhat{e}_y(\phi), \vhat{e}_0(\phi)\right)$, aligned with this sublattice spin direction characterized by $\phi$. Here, $\vhat{e}_0(\phi)$ points in the direction of the sublattice spin in the ground state and $\vhat{e}_x(\phi), \vhat{e}_y(\phi) $ are two mutually perpendicular directions to $\vhat{e}_0(\phi)$.
We further define, 
\begin{equation}
\vhat{e}_{\pm}(\phi) \equiv \left(\vhat{e}_x(\phi) \pm i \,\vhat{e}_y(\phi)\right)/\sqrt{2}.
\label{app.local-frame}
\end{equation}
We then define the \emph{local exchanges} as 
\begin{equation}
\mathcal{J}^{\mu\nu}_{\vec{\gamma}}(\phi)\equiv \trp{\vhat{e}}_{\mu}(\phi) \mat{J}_{\vec{\gamma}} \vhat{e}_{\nu}(\phi),
\label{eq.loc-exchange}
\end{equation}
where $\vhat{e}_\mu(\phi), \vhat{e}_\nu(\phi)$ are $\vhat{e}_{+}(\phi), \vhat{e}_{-}(\phi),$ and $\vhat{e}_0(\phi)$, and $\vec{\gamma} = \pm\vec{x}, \pm\vec{y}$, the nearest-neighbor bonds. 
The Fourier transforms of the exchange matrix elements, $\mathcal{J}^{\mu\nu}_{\vec{\gamma}}(\phi)$,
are defined as 
\begin{equation}
\mathcal{J}^{\mu\nu}_{\vec{q}}(\phi) \equiv \sum_{\vec{\gamma}}\exp{(-i\vec{q}\cdot\vec{\gamma})} \mathcal{J}^{\mu\nu}_{\vec{\gamma}}(\phi). 
\label{eq.exchange-FT}
\end{equation}
Performing the Holstein-Primakoff expansion~\cite{Auerbach1998} to $O(S)$ on this model yields~\cite{Rau2018,Khatua2023}
\begin{equation}
{\cal H} \approx NS(S+1)\epsilon + {\cal H}_2,
\label{H2-H4}
\end{equation}
where $\epsilon = \frac{1}{2}\mathcal{J}^{00}_{\vec{q}=\vec{0}}(\phi)$ and
\begin{equation}
\label{eq.h}
  {\cal H}_2 = \sum_{\vec{q}} \left[
                {A}^{\phantom\dagger}_{\vec{q}}(\phi)\,\h{a}_{\vec{q}}\l{a}_{\vec{q}} +
          \frac{1}{2!} \left(
          {B}^{\phantom\dagger}_{\vec{q}}(\phi)\,\h{a}_{\vec{q}}\h{a}_{-\vec{q}} +
          \cc{B}_{\vec{q}}(\phi)\,\l{a}_{-\vec{q}}\l{a}_{\vec{q}}\right)
          \right],
\end{equation}
 with $\h{a}_{\vec{q}}$ ($\l{a}_{\vec{q}}$) is the bosonic creation (annihilation) operator at wave vector $\vec{q}$. 
 Here ${\cal H}_2$ denotes the linear spin-wave Hamiltonian. 
 In terms of the local exchanges, we have
\begin{align}
A_{\vec{q}}(\phi) &= S\left(\mathcal{J}^{+-}_{\vec{q}}(\phi) - \mathcal{J}^{00}_{\vec{0}}(\phi)\right),\nonumber \\
B_{\vec{q}}(\phi) &= S \mathcal{J}^{++}_{\vec{q}}(\phi).
    \label{eq.HP-coeff}
\end{align}

\noindent This linear spin-wave Hamiltonian [Eq.~\eqref{eq.h}] can be diagonalized using a Bogoliubov transformation~\cite{Auerbach1998}. Defining the matrix 
\begin{equation}
    \mat{M}_{\vec{q}}(\phi) \equiv \left(\begin{array}{cc}
        A_{\vec{q}}(\phi) &  B_{\vec{q}}(\phi)\\
        \cc{B}_{\vec{q}}(\phi) & A_{\vec{q}}(\phi)
    \end{array}\right),
\end{equation}
the linear spin-wave energy spectrum is given by the eigenvalues of $\mat{\sigma}_z \mat{M}_{\vec{q}}(\phi)$, where $\mat{\sigma}_z$ is a (block) Pauli matrix and the spectrum is
\begin{equation}
\omega_{\vec{q}}(\phi) = \sqrt{A_{\vec{q}}(\phi)^2 - |{B}_{\vec{q}}(\phi)|^2}.
\label{eq.lsw-spectrum}
\end{equation}
We next derive the Fourier transformed local exchanges [Eq.~\eqref{eq.exchange-FT}] in different regimes of the phase angle, from which we can calculate the linear spin-wave spectrum using Eq.~\eqref{eq.HP-coeff} and Eq.~\eqref{eq.lsw-spectrum}. 
\subsection{Regime-I}
In this regime, the classical ground state is a N\'eel state with two sublattices given by,
\begin{eqnarray}
\vec{S}_A = +S(\cos{\phi}\,\vhat{x}+\sin{\phi}\,\vhat{y}),\nonumber\\
\vec{S}_B = -S(\cos{\phi}\,\vhat{x}+\sin{\phi}\,\vhat{y}).
\label{eq.neel-sublat}
\end{eqnarray}
Note that $\{\vhat{x},\vhat{y},\vhat{z}\}$ is the \emph{global} coordinate frame in the spin space. We now consider a canonical transformation, $\pi$-rotation of the spins on one of the two sublattices of the square lattice about the $\vhat{z}$ axis. This transformation changes the Hamiltonian as well as the ground state configuration. The change in the Hamiltonian can be expressed as a change in the coupling exchange matrix of the Hamiltonian [Eq.~\eqref{eq.Ham1}],
\begin{equation}
    \mat{J}_{\vec{x}} = \mat{J}_{-\vec{x}}= \left(\begin{array}{ccc}
        J+K &  0 &0\\
        0 & J & 0\\
        0& 0& J
    \end{array}\right) \rightarrow\mat{\tilde{J}}_{\vec{x}} = \mat{\tilde{J}}_{-\vec{x}}= \left(\begin{array}{ccc}
        -J-K &  0 &0\\
        0 & -J & 0\\
        0& 0& J
    \end{array}\right),\nonumber 
\end{equation}
\begin{equation}
    \mat{J}_{\vec{y}} = \mat{J}_{-\vec{y}}= \left(\begin{array}{ccc}
        J &  0 &0\\
        0 & J+K & 0\\
        0& 0& J
    \end{array}\right) \rightarrow\mat{\tilde{J}}_{\vec{y}} = \mat{\tilde{J}}_{-\vec{y}}= \left(\begin{array}{ccc}
        -J &  0 &0\\
        0 & -J-K & 0\\
        0& 0& J
    \end{array}\right).
    \label{eq.exchange-rI}
\end{equation}
By this transformation, the N\'eel ground state configuration changes to a ferromagnetic configuration in the $\vhat{x}-\vhat{y}$ plane which has only one sublattice given by,
\begin{eqnarray}
\vec{S}_{\vec{r}} = S(\cos{\phi}\,\vhat{x}+\sin{\phi}\,\vhat{y}).
\label{eq.app-ferro}
\end{eqnarray}
The advantage of performing this transformation is that we now have only one sublattice describing the ground state, making the one-sublattice spin-wave analysis discussed above directly applicable in Regime-I to find the spin-wave spectrum [Eq.~\eqref{eq.lsw-spectrum}]. 

We now define a local frame aligned with an arbitrary ferromagnetic ground state parameterized by an angle $\phi$ [Eq.~\eqref{eq.app-ferro}],
\begin{subequations}
\label{eq.loc-frame-ferro}
\begin{align}
\vhat{e}_x(\phi) &= -\sin{\phi}\,\vhat{x}+\cos{\phi}\,\vhat{y}, \\
\vhat{e}_y(\phi) &= \vhat{z}, \\
\vhat{e}_0(\phi) &= \cos{\phi}\,\vhat{x}+\sin{\phi}\,\vhat{y},
\end{align}
\end{subequations}
and have the corresponding $\vhat{e}_{\pm}(\phi)$ as defined in Eq.~\eqref{app.local-frame}.
We then define the \emph {local exchanges} as done for Eq.~\eqref{eq.loc-exchange}$, \mathcal{J}^{\mu\nu}_{\delta}(\phi)= \trp{\vhat{e}}_{\mu}(\phi) \mat{\tilde{J}}_{\delta} \vhat{e}_{\nu}(\phi)$.
Using Eq.~\eqref{eq.exchange-FT}, we obtain the Fourier transform of the local exchanges, which are necessary to compute the linear spin-wave spectrum,
\begin{align*}
    \mathcal{J}^{+-}_{\vec{q}}(\phi) &= -K\left({\rm sin}^2\phi \cos{q_x} + {\rm cos}^2\phi \cos{q_y}\right), \\
    \mathcal{J}^{00}_{\vec{q}}(\phi) &= 
    -2\left(J+K{\rm cos}^2\phi \right)\cos{q_x}- 2\left(J+K{\rm sin}^2\phi \right)\cos{q_y},  \\
  \mathcal{J}^{++}_{\vec{q}}(\phi) &= -\left(2J+K{\rm sin}^2\phi \right)\cos{q_x}- \left(2J+K{\rm cos}^2\phi \right)\cos{q_y}.
\end{align*}
Note that $\mathcal{J}^{00}_0(\phi) = -2(2J+K)$. 
Using these Fourier transformed local exchanges, we can compute the linear spin wave spectrum, $\omega_{\vec{q}}(\phi)$, using Eq.~\eqref{eq.lsw-spectrum} as a function of $\phi$. 
With this spectrum in hand, we next compute the zero-point energy, $\epsilon_{\rm{Q}}(\phi) = (1/2)\sum_{\vec{q}}\omega_{\vec{q}}(\phi)$ as a function of $\phi$. 
The zero-point energy is found to have minima at $\phi =0, \pi/2, \pi, 3\pi/2$, which correspond to the N\'eel states along $\pm\vhat{x},\pm\vhat{y}$ directions and these are thus the states picked by quantum ObD at zero temperature in Regime-I.

\subsection{Regime-IV}
In this regime, the classical ground states are ferromagnetic states pointing along arbitrary directions in the $\vhat{x}-\vhat{y}$ plane. To perform the spin-wave analysis, we start with the same reference ground state as in Eq.~\eqref{eq.app-ferro} and the same local frame convention as in Eq.~\eqref{eq.loc-frame-ferro}. Using the coupling exchange matrix in the global frame in this regime, given by Eq.~\eqref{eq.exchange-global}, we obtain the following Fourier transformed local exchanges  
\begin{align*}
    \mathcal{J}^{+-}_{\vec{q}}(\phi) &= \left(2J+K{\rm sin}^2\phi \right)\cos{q_x} + \left(2J+K{\rm cos}^2\phi \right)\cos{q_y}, \\
    \mathcal{J}^{00}_{\vec{q}}(\phi) &= 
    2\left(J+K{\rm cos}^2\phi \right)\cos{q_x}+2\left(J+K{\rm sin}^2\phi \right)\cos{q_y},  \\
  \mathcal{J}^{++}_{\vec{q}}(\phi) &= \left(K {\rm sin}^2\phi\right)\cos{q_x}+\left(K {\rm cos}^2\phi\right)\cos{q_y}.
\end{align*}
Using these exchanges, we compute the spectrum using Eq.~\eqref{eq.lsw-spectrum}, $\omega_{\vec{q}}(\phi)$, as a function of $\phi$. 
The zero-point energy is found to have minima at $\phi = 0, \pi/2,\pi,3\pi/2$, corresponding to the ferromagnetic states along $\pm\vhat{x},\pm\vhat{y}$ directions, which result from quantum ObD at zero temperature.

\bibliography{main}

\end{document}